\documentclass[twocolumn]{aastex701}

\usepackage{amsmath}
\usepackage{xcolor}

\newcommand{\cii}{[\ion{C}{2}]}
\newcommand{\Msun}{M_\odot}
\newcommand{\Lsun}{L_\odot}
\newcommand{\Tb}{T_b}
\newcommand{\beff}{b_{\rm eff}}
\newcommand{\Pshot}{P_{\rm shot}}
\newcommand{\Ptwohalo}{P_{\rm 2h}}
\newcommand{\Mone}{M_1}
\newcommand{\kc}{k_c}
\newcommand{\Rpf}{A_{\rm pf}}
\newcommand{\hMpc}{h\,{\rm Mpc}^{-1}}
\newcommand{\PkUnit}{h^{-3}\,{\rm Mpc}^{3}\,\mu{\rm K}^{2}}
\newcommand{\LCII}{L_{\rm [CII]}}
\newcommand{\Mh}{M_h}
\newcommand{\MUV}{M_{\rm UV}}

\begin{document}

\title{An Empirically Calibrated [C\,II] Luminosity--Halo Mass Relation for Line Intensity Mapping}

\author[0000-0001-5893-7697]{Oliver Jeong}
\affiliation{CNRS-UCB International Research Laboratory, Centre Pierre Bin\'etruy, IRL 2007, CPB-IN2P3, Berkeley, CA 94720, USA}
\affiliation{Lawrence Berkeley National Laboratory, 1 Cyclotron Road, Berkeley, CA 94720, USA}
\email[show]{objeong@lbl.gov}

\author[0000-0001-6320-261X]{Yi-Kuan Chiang}
\affiliation{Academia Sinica Institute of Astronomy and Astrophysics (ASIAA), Taipei 106319, Taiwan}
\email{ykchiang@asiaa.sinica.edu.tw}

\author[0000-0001-9912-5070]{Martin White}
\affiliation{Lawrence Berkeley National Laboratory, 1 Cyclotron Road, Berkeley, CA 94720, USA}
\affiliation{Berkeley Center for Cosmological Physics, Department of Physics, UC Berkeley, CA 94720, USA}
\affiliation{Department of Physics, University of California, Berkeley, CA 94720}
\email{mwhite@berkeley.edu}

\begin{abstract}
Millimeter-wave line intensity mapping (LIM) aims to observe the high-redshift universe through the 158\,$\mu$m fine-structure line of \cii, the brightest far-infrared line of star-forming galaxies. Whether the clustering of large-scale structure, which carries rich cosmological information, is measurable in the \cii\ signal at $z>4$ or buried under the shot noise of the emitters is determined from the amplitude and shape of the \cii\ power spectrum. We calibrate $L(\Mh,z)$ on \cii\ data alone: the mean \cii\ intensity from tomographic cross-correlation at $0<z<4.2$ and the luminosity-function measurements and limits at $z\simeq0$ and $4.3\lesssim z\lesssim8$. We model the relation as a double power law and fit its four parameters at $z=4.5$ and 5.5 separately, with no assumption about their evolution, and then jointly across $0<z<7$, and we examine how the uncertainty depends on the modeling assumptions. The shape of the power spectrum is settled by the existing data. The clustering peak stands a factor of $45\,(23$--$93)$ and $64\,(38$--$112)$ above the shot-noise floor at $z=4.5$ and 5.5, and the two cross at $\kc\simeq0.37$ and $0.44\,\hMpc$. The amplitude is known less well, the clustering power to 0.5--0.6\,dex and the shot noise to 0.7--0.9\,dex. The luminosity-weighted bias, which together with the mean intensity sets the clustering amplitude, is known to about 10\% within the assumptions of this work and is relatively high, about 4. The results show that the signal comes from halos of $10^{11}$--$10^{12}\,\Msun$, similar in mass to the hosts of the Lyman-break galaxies and Ly$\alpha$ emitters that galaxy surveys map at these redshifts.
\end{abstract}

\section{Introduction} \label{sec:intro}

Line intensity mapping measures the aggregate emission of a spectral line from all galaxies in a cosmological volume, without resolving them \citep{Bernal2022, Lidz2026}. The 158\,$\mu$m fine-structure line of singly ionized carbon, \cii, is the brightest far-infrared cooling line of star-forming galaxies and redshifts into the millimeter and submillimeter atmospheric windows at $z\gtrsim4$. This makes it the primary target of ground-based LIM experiments that aim to probe the high-redshift universe and the scientific gains that come with it \citep{Karkare2022, Sailer2021, Ferraro2022}, among them primordial non-Gaussianity \citep{MoradinezhadDizgah2019}, the dark energy equation of state \citep{MoradinezhadDizgah2024}, and the sum of the neutrino masses \citep{MoradinezhadDizgah2022}. The same maps carry astrophysical information, as they measure the \cii\ emission of all galaxies at a redshift rather than of those bright enough to be selected, and with it the star formation and the interstellar medium of the galaxy population at $z>4$ \citep{Kovetz2017}. It is the target of several current-generation experiments, including TIME \citep{Sun2021} and FYST \citep{CCAT2023}, and several spectrometer technologies have been developed to observe it \citep{Jeong2026, DESHIMA, DESHIMA2, SuperSpec_redford_2022}. Optical galaxy surveys at $z\gtrsim4$ become increasingly expensive due to the faintness of the sources, and \cii\ offers a distinct opportunity to probe high-redshift large-scale structure over large volumes. However, the measurement is also challenging. Continuum foregrounds, lines from other redshifts that fall into the same band, and the stability of the atmosphere all stand between the maps and the power spectrum \citep{Bernal2022, Lidz2026}, and none of them can be assessed without a model of the signal. This opportunity motivates an empirically calibrated model of \cii\ emission at $z\gtrsim4$.

The main observable of a LIM survey is the power spectrum of its map, the variance of the intensity as a function of spatial scale. On large scales it follows the clustering of the halos that host the emitters, scaled by the mean intensity and by the bias of those halos, and on small scales it flattens to the shot noise of the discrete emitters. Whether such an experiment detects the clustering signal, and what cosmological information it then recovers, depends on both parts. The abundance and clustering of dark matter halos are well characterized, so what a model needs is a way to connect the line emission to the halos. The most common way is a relation between the mass of the halo that hosts the emitters and the \cii\ luminosity, $L(\Mh,z)$, which fixes the mean brightness temperature $\Tb$, the luminosity-weighted bias, and the shot-noise power. The $L(\Mh)$ relations commonly adopted as fiducial fall into two families.

The first family chains empirical scaling relations. A halo is assigned a star formation rate (SFR) through abundance matching or a fitted SFR--$\Mh$ relation, and the SFR is converted to $\LCII$ with a relation calibrated mostly on local galaxies \citep[e.g.,][]{Gong2012, Silva2015, Yue2015, Serra2016, Chung2020, Bethermin2022}. \citet{Silva2015}, for example, take the SFR of each halo from a semi-analytic model of galaxy formation and convert it with an $\LCII$--SFR relation built from the \cii-to-far-infrared ratio of nearby late-type galaxies, to predict the signal at $z=5$--8. The challenge of this approach is that each link is measured in a regime far from the one where it is applied, and the errors of the links are difficult to propagate to the predicted power spectrum. Predictions from different chains differ by more than an order of magnitude at the same redshift \citep{Chung2020, Bernal2022}.

The second family models $L(\Mh)$ from simulations. Mock galaxies from semi-analytic or hydrodynamic models are painted with a model of the line emission of their interstellar medium \citep[e.g.,][]{Lagache2018, Popping2019, Leung2020, Vizgan2022, Yang2022}. The \cii\ luminosity of a galaxy is set by the density, metallicity, and radiation field of its gas on the scale of molecular clouds, far below the resolution of any cosmological simulation. Each model therefore prescribes these conditions, assuming how the gas is arranged into clouds and how strongly it is illuminated, and its predicted luminosities carry those assumptions \citep{Popping2019}. At $z>4$ the only data that could test them are the few existing \cii\ detections, so the uncertainty of a simulated $L(\Mh)$ is no easier to quantify than that of a chained one.

A third approach, and the one closest to this work, fits a parametric $L(\Mh,z)$ directly to \cii\ data. \citet{Padmanabhan2019} constrained it with the local LF, the cosmic star formation history, and the Planck cross-correlation measurement of the \cii\ intensity at $z\simeq2.6$ \citep{Pullen2018}, with the high-redshift limits then available as consistency checks. The work presented in this paper takes this empirical approach with the updated \cii\ data available today: $L(\Mh,z)$ is constrained by \cii\ data alone, and the uncertainty of the data is propagated to the power spectrum. This work sets out to answer a crucial question from the cosmologist's perspective: whether the clustering of large-scale structure in the \cii\ signal at $z>4$ is within reach of a LIM experiment or buried under the shot noise, and on which scales it is detectable. The answer rests on two properties of the power spectrum. Its amplitude sets the signal-to-noise ratio of a survey, and its shape decides on which scales the clustering stands above the shot-noise floor. The core question of this work is the degree to which the existing \cii\ data can already inform the design of a \cii\ LIM experiment, through the shape and the amplitude of the power spectrum.

A recent measurement of the mean \cii\ intensity provides an important new constraint. \citet{Chiang2026} detected it over $0<z<4.2$ by tomographic cross-correlation of broadband far-infrared maps with spectroscopic reference samples, which pins down the first moment of $L(\Mh)$ over the halo mass function (HMF). Observations with ALMA have accumulated luminosity function (LF) measurements and limits at $z\simeq4.3$--8, from targeted \citep{Yan2020}, serendipitous \citep{Loiacono2021}, and blind \citep{Matsuda2015, Yamaguchi2017, Uzgil2021} searches, and the local LF \citep{Hemmati2017} anchors the relation at $z=0$. The available data have changed in two ways since the calibration of \citet{Padmanabhan2019}. At $z\simeq2.6$ the new mean intensity lies more than an order of magnitude below the value of \citet{Pullen2018} used there, a difference caused by correlated continuum emission of the cosmic infrared background that contaminated the earlier measurement \citep{Chiang2026}, and most of the ALMA constraints on the LF have been published since. These changes motivate a new calibration of $L(\Mh,z)$. We fit a double power-law $L(\Mh,z)$ to these data, first at single redshifts, where the four parameters are constants and nothing is assumed about their evolution, and then jointly over $0\le z\le7$ with the minimal set of assumptions that lets the parameters evolve, a single power law in $(1+z)$ per parameter. No scaling relation measured in another regime enters either fit, and the full posterior is propagated to the power spectrum, so the error bars of the LIM observables follow from the data. The result is meant as a reference for the design and forecasts of \cii\ LIM experiments.

Section~\ref{sec:methods} describes the model, the data, and the inference. Section~\ref{sec:results} presents the results and what they imply for LIM measurements: fits of the four parameters of $L(\Mh)$ independently at $z=4.5$ and $z=5.5$, the redshifts where LF data exist and where ground-based \cii\ LIM operates (Section~\ref{sec:singlez}), their power spectra with the decomposition by halo mass (Section~\ref{sec:pk}), a joint fit of all redshifts with the simplest redshift evolution (Section~\ref{sec:global}), and the implications for a LIM experiment and the measurement that would move the calibration forward (Section~\ref{sec:implications}). Section~\ref{sec:summary} concludes.

Throughout we assume a flat $\Lambda$CDM cosmology with $h=0.6766$, $\Omega_m=0.3111$, $\Omega_b=0.0490$, $n_s=0.9665$, and $A_s=2.105\times10^{-9}$ \citep{Planck2018}. Unless stated otherwise, quoted values are posterior medians with central 68\,\% intervals.

\section{Methods} \label{sec:methods}

\subsection{Model} \label{sec:model}

The model is built on the halo population. It assumes that the \cii\ emission comes from halos and that the luminosity is solely a property of the halo mass at the time of emission. The luminosity of a galaxy could also depend on its environment, its recent mergers, or other properties, and the assumption is the minimal one. It is common in the LIM literature, and it builds on the well-understood abundance and clustering of dark matter halos over cosmic time. The halo model itself is an approximation, suited to qualitative and order-of-magnitude statements, which is the level the present data support. Each halo is treated as a single emitter at its center, which is adequate on the clustering scales, $k\lesssim1\,\hMpc$, where a halo is unresolved.

\subsubsection{Halo mass--luminosity relation}

The median \cii\ luminosity of a halo of mass $\Mh$ is taken to be a double power law in the mass of the hosting halo at the time of emission,
\begin{equation} \label{eq:LM}
  \frac{L(\Mh)}{\Lsun} = 2N\,\frac{\Mh}{\Msun}
  \left[\left(\frac{\Mh}{\Mone}\right)^{-\alpha}
       +\left(\frac{\Mh}{\Mone}\right)^{\beta}\right]^{-1}.
\end{equation}
The break mass $\Mone$ separates a low-mass regime, $L\propto \Mh^{1+\alpha}$, from a high-mass regime, $L\propto \Mh^{1-\beta}$, and $N$ sets the light-to-mass ratio at the break, $L(\Mone)/\Mone=N$. The slope $\alpha$ controls how sharply emission switches off below $\Mone$, and therefore the faint end of the LF and the bias of the emitters. The slope $\beta$ controls how quickly emission saturates in massive halos, and therefore the bright end and the shot noise.

Individual halos are assumed to scatter about the median with a log-normal distribution of width $\sigma$. The data used here cannot determine $\sigma$: it enters $\Tb$ only through a factor that is degenerate with $N$ (Equation~\ref{eq:Tb}), so it is held fixed at a value taken from the literature, $\sigma=0.4$\,dex. This is the quadrature sum of the dispersion of the ALPINE $\LCII$--SFR relation, 0.28\,dex \citep{Schaerer2020}, and of the dispersion in SFR at fixed halo mass, about 0.3\,dex in empirical models of galaxy growth \citep{Behroozi2019}. The semi-analytic model of \citet{Yang2022} gives 0.34\,dex at $z=1$ falling to 0.2\,dex at $z\ge5$. Appendix~\ref{app:sigma} shows how the power spectrum changes when $\sigma$ is set to 0.2 or 0.6\,dex instead.

\subsubsection{From $L(\Mh)$ to observables} \label{sec:observables}

Under these assumptions the mean brightness temperature is given by
\begin{equation} \label{eq:Tb}
  \Tb(z) = C_{LT}(z)\,e^{(\sigma\ln10)^2/2}\int d\Mh\,\frac{dn}{d\Mh}\,L(\Mh,z),
\end{equation}
with
\begin{equation}
  C_{LT}(z) = \frac{c^3(1+z)^2}{8\pi k_B\,\nu_{\rm rest}^3\,H(z)},
\end{equation}
where $\nu_{\rm rest}=1900.54$\,GHz. $dn/d\Mh$ is the HMF of \citet{ShethTormen1999}, integrated over $10^{9}$--$10^{15}\,\Msun$, and the halo bias is from the same model. The exponential converts the median of the log-normal to its mean. The LF follows from convolving the HMF with the scatter,
\begin{equation} \label{eq:LF}
\begin{split}
  \phi(L) &\equiv \frac{dn}{d\log_{10}L}
  = \int d\Mh\,\frac{dn}{d\Mh}\,\frac{1}{\sqrt{2\pi}\,\sigma}\\
  &\quad\times\exp\!\left[-\frac{\big(\log_{10}L-\log_{10}L(\Mh)\big)^2}{2\sigma^2}\right].
\end{split}
\end{equation}

The anisotropy of the LIM power spectrum contains significant information on cosmology \citep{SchaanWhite_2021_1, SchaanWhite_2021_2, Bernal2022, Lidz2026}, but for detectability the angle-averaged quantity, the power spectrum monopole, is the simplest to consider. The large-scale biases inferred below are high, which suppresses the anisotropy from redshift-space distortions, and the rest of the paper uses the monopole. The power spectrum is then the sum of a clustering (two-halo) term and the shot noise of the emitters \citep{SchaanWhite_2021_2},
\begin{equation} \label{eq:Pk}
  P(k,z) = \left\langle D_{\rm RSD}\,\Ptwohalo(k,z)\right\rangle_\mu + \Pshot(z) ,
\end{equation}
with
\begin{align}
  \Ptwohalo(k,z) &= \Tb^2\,\beff^2\,P_m(k,z), \label{eq:P2h}\\
  \Pshot(z) &= C_{LT}^2\,e^{2(\sigma\ln10)^2}\int d\Mh\,\frac{dn}{d\Mh}\,L^2(\Mh,z) , \label{eq:Pshot}
\end{align}
where $P_m$ is the linear matter power spectrum from CAMB \citep{Lewis2000} and the luminosity-weighted bias is
\begin{equation} \label{eq:beff}
  \beff(z) = \frac{\int d\Mh\,(dn/d\Mh)\,L(\Mh,z)\,b(\Mh,z)}{\int d\Mh\,(dn/d\Mh)\,L(\Mh,z)},
\end{equation}
with $b(\Mh,z)$ the scale-independent halo bias of \citet{ShethTormen1999}. The linear $P_m$ suffices on the scales considered here. At $z=4.5$ and 5.5, the non-linear spectrum of \citet{Takahashi2012} exceeds it by less than 1\% at $k<0.1\,\hMpc$, which leaves the peak of the two-halo term unchanged, and by 6--7\% at $k=0.4\,\hMpc$, roughly beyond which the shot noise dominates. Each halo is a single emitter, so its Poisson term is the shot noise and there is no separate one-halo term. A model in which several galaxies occupy a halo would carry a halo term as well as the galaxy shot noise, set by the conditional LF \citep[Appendix A of][]{SchaanWhite_2021_2}, and the two coincide when every halo holds one emitter. The extent of the emission within a halo is not modeled. Redshift-space distortions enter through the factor $D_{\rm RSD}=(1+f\mu^2/\beff)^2\,[1+(k\mu\sigma_v)^2/2]^{-2}$, which combines the enhancement from supercluster infall \citep{Kaiser1987} with a Lorentzian damping for the fingers of god. Here $f$ is the scale-independent linear growth rate, $\sigma_v$ an effective velocity dispersion, and $\langle\cdot\rangle_\mu$ the average over the cosine $\mu$ of the angle to the line of sight, which gives the monopole. All halo-model quantities are computed with SSLimPy\footnote{\url{https://github.com/Sefa76/SSLimPy}} \citep{SSLimPy}.

Two derived quantities describe the shape of the power spectrum and how it informs the design of a LIM experiment. The floor is the shot noise, the part of the power that does not trace large-scale structure. With the monopole of the two-halo term defined as $P^{(0)}_{\rm 2h}\equiv\langle D_{\rm RSD}\,\Ptwohalo\rangle_\mu$,
\begin{equation}
  \Rpf \equiv \left.\frac{P^{(0)}_{\rm 2h}}{\Pshot}\right|_{k_{\rm peak}}
\end{equation}
is the height of the two-halo term above the floor at the wavenumber $k_{\rm peak}$ where the monopole is largest, and $\kc$ is defined by
\begin{equation}
  P^{(0)}_{\rm 2h}(\kc) = \Pshot ,
\end{equation}
the wavenumber at which the two-halo term equals the floor. A large $\Rpf$ means that a measurement at $k<\kc$ is dominated by the clustering of large-scale structure, so it is sensitive to $\Tb\beff$, to the shape of $P_m$, to redshift-space distortions, and to any scale dependence of the bias on large scales from primordial non-Gaussianity. A small $\Rpf$ means that it mostly measures the second moment of the LF, which is set by the astrophysics of the line emission.

\subsubsection{Redshift evolution} \label{sec:evolution}

For fits at a single redshift the four parameters $\{\log_{10}\Mone,\log_{10}N,\alpha,\beta\}$ are constants. For the joint fit of all redshifts each parameter follows a single power law in $(1+z)$,
\begin{equation} \label{eq:zevol}
\begin{split}
  \log_{10}X(z) &= X_{p1} + X_{p2}\,u, \qquad X\in\{\Mone,N\},\\
  Y(z) &= Y_{p1}\,10^{\,Y_{p2}\,u}, \qquad\;\; Y\in\{\alpha,\beta\},
\end{split}
\end{equation}
with $u\equiv\log_{10}[(1+z)/(1+z_p)]$ and pivot $z_p=2$, where the brightness temperature measurement of \citet{Chiang2026} is most precise. The first coefficient is then the value at $z=2$ and the second the logarithmic rate of change, $\Mone,N\propto(1+z)^{X_{p2}}$ and $\alpha,\beta\propto(1+z)^{Y_{p2}}$. This is the simplest form that lets all four parameters evolve, and it gives eight coefficients in total. A power law cannot change sign, so $\alpha$ and $\beta$ stay positive at all redshifts. There is no compelling reason for the parameters to evolve in this simple way, but the data do not constrain a more complex form. Section~\ref{sec:global} compares the joint fit with the single-redshift fits to show how this assumption affects the inference.

\subsection{Data} \label{sec:data}

Table~\ref{tab:data} lists the data sets that constrain the model and how each enters the likelihood (Section~\ref{sec:inference}). Each is described in turn.

\begin{deluxetable*}{llccl}
\tablecaption{Data used in the calibration \label{tab:data}}
\tablehead{
\colhead{Data set} & \colhead{Ref.} & \colhead{Redshift} &
\colhead{$N_{\rm data}$} & \colhead{Enters as}
}
\startdata
$\Tb(z)$ curve & 1 & 0--4.2 & 2 & measurement, rank-two covariance \\
Local LF & 2 & 0.02 & 12 & measurements \\
ALPINE targeted LF & 3 & 4.5, 5.5 & 7 & lower limits \\
ALPINE serendipitous LF & 4 & 4.3--5.4 & 3 & upper limits \\
Blind searches & 5, 6, 7 & 4.3--8 & 12 & upper limits on counts\tablenotemark{a}
\enddata
\tablerefs{(1) \citet{Chiang2026}; (2) \citet{Hemmati2017}; (3) \citet{Yan2020}; (4) \citet{Loiacono2021}; (5) \citet{Matsuda2015}; (6) \citet{Yamaguchi2017}; (7) \citet{Uzgil2021}.}
\tablenotetext{a}{Included in the likelihood but not approached by any model in the posteriors.}
\end{deluxetable*}

\subsubsection{Mean \cii\ brightness temperature} \label{sec:data_tb}

\citet{Chiang2026} detected the cosmic \cii\ background at $3\sigma$ by cross-correlating broadband intensity maps from Planck, Herschel, and IRAS with spectroscopic galaxies and quasars from the Sloan Digital Sky Survey. \citet{Chiang2026} reports the \cii\ luminosity density as
\begin{equation}
  \rho_{\rm [CII]}(z) = \left[R_0 + C\log_{10}(1+z)\right]
  \nu_{\rm [CII]}\,\epsilon_{\rm cont}(\nu_{\rm [CII]},z),
\end{equation}
where $\epsilon_{\rm cont}$ is the continuum emissivity at the line frequency and $R_0$ and $C$ describe the line-to-continuum ratio. $\rho_{\rm [CII]}$ is converted to $\Tb$ with $C_{LT}$. Because the published redshift dependence is a two-parameter family, the $\Tb$ values at different redshifts are almost fully correlated. The published posterior chain is propagated to a mean and a rank-two covariance matrix for $\Tb$ at the bin centers, which run from $z=0$ to 7 in steps of 0.5, and its pseudo-inverse is used in the likelihood.

The $\Tb$ measurement ends at $z=4.2$. At $z=4.5$ and 5.5, where the ALMA LF data lie, an extrapolation of the two-parameter form serves as the constraint on $\Tb$. The evolution up to $z=4.2$ is smooth, which makes the extrapolation plausible, and it is the only information on the first moment of $L(\Mh)$ at those redshifts.

\subsubsection{Luminosity functions} \label{sec:data_lf}

At $z\simeq0$ the 12 bins of the local \cii\ LF of \citet{Hemmati2017} enter as measurements.

At $z=4.5$ and 5.5 the LF comes from the ALPINE targeted detections \citep{LeFevre2020, Yan2020}. ALPINE targets were selected in the rest-frame UV, so \cii\ emitters that are faint in the UV are missing, and the targeted LF is a lower bound on the true one. Each ALPINE bin therefore enters every fit as a lower limit, placed conservatively one error bar below the published value.

The serendipitous \cii\ emitters in the ALPINE fields \citep{Loiacono2021} are free of the UV selection, but they are found in the volumes around the targets, which are overdense, and their LF therefore overestimates the field value. It enters as an upper limit. The blind searches of \citet{Matsuda2015}, \citet{Yamaguchi2017}, and \citet{Uzgil2021} provide upper limits on cumulative counts at $4.3<z<8$ from zero detections. They are included in the likelihood, but no model in the posteriors comes close to any of them, so they do not affect the results and are not shown in the figures.

The limits rest on small volumes. The targeted LF is normalized by the UV LF of the ALPINE parent samples over about 2\,deg$^2$ of COSMOS and ECDFS in windows of $\Delta z\simeq0.2$--0.7, the serendipitous LF comes from the ALPINE cubes themselves, and the blind searches cover a few hundred arcmin$^2$ at most. The UV selection acts in one direction and is the reason the targeted LF enters as a lower limit. Cosmic variance acts in both directions: a limit measured in an overdense field is too high and one measured in an underdense field is too low. Each limit therefore receives a cosmic-variance width, the rms fluctuation of the count in the survey window computed from the linear power spectrum and the bias of the sample. It is $\sigma_{\rm CV}=6$--10\,\% for the ALPINE bins, 82\,\% for the serendipitous LF, whose volume is a few thousand Mpc$^3$, 25--48\,\% for the blind searches, and 3\,\% for the local LF. This width is added in quadrature to the published error, and every limit, including the zero-detection limits, is placed one error bar beyond the published value as a conservative measure, as for the ALPINE bins, which corresponds to 84\,\% confidence.

\subsection{Inference} \label{sec:inference}

\subsubsection{Likelihood}

The likelihood follows $\mathcal{L}\propto e^{-\chi^2/2}$ with
\begin{equation}
\begin{split}
  \chi^2 = {}& \mathbf{r}^{T}\mathbf{C}^{+}\mathbf{r}
         + \sum_{i}\frac{\big[\phi_i^{\rm obs}-\phi(L_i)\big]^2}{\sigma_i^2}\\
         & + A\sum_{l}\max(0,\Delta_l)^2 .
\end{split}
\end{equation}
The first two terms are the measurements. $\mathbf{r}$ is the residual vector of $\Tb$ at the bin centers and $\mathbf{C}^{+}$ the pseudo-inverse of its covariance, and the sum runs over the 12 bins of the local LF. The last term is the upper and lower limits. $\Delta_l$ is the amount in dex by which the model crosses limit $l$, so the term vanishes when every limit is respected and, with $A=10^5$, acts as a wall once one is crossed. The posterior is this likelihood times flat priors within the ranges of Table~\ref{tab:priors}, and the quoted results are its medians and intervals.

\subsubsection{Sampling and priors}

The best fit is located by differential evolution (\texttt{scipy.optimize.differential\_evolution}, four independent seeds) and refined with a Nelder--Mead search, and the posterior is sampled with the affine-invariant ensemble sampler \texttt{emcee} \citep{ForemanMackey2013}, using 40 walkers and $3\times10^5$ steps for the joint fit and $4\times10^4$ for each single-redshift fit. The chains span 76 autocorrelation times for the joint fit and more than 150 for the single-redshift fits. Priors are flat within the ranges of Table~\ref{tab:priors}, and both exponents are bounded at every fitted redshift, not only at the pivot.

\begin{deluxetable}{lccc}
\tablecaption{Priors \label{tab:priors}}
\tablehead{\colhead{Parameter} & \colhead{Single redshift} & \colhead{Joint, $p1$} & \colhead{Joint, $p2$}}
\startdata
$\log_{10}(\Mone/\Msun)$ & $[10,\,13]$ & $[10,\,13]$ & $[-3,\,3]$ \\
$\log_{10}N$ & $[-8,\,-2]$ & $[-8,\,-2]$ & $[-6,\,6]$ \\
$\alpha$ & $[0,\,4]$ & $[0,\,4]$ & $[-0.96,\,0.96]$\tablenotemark{a} \\
$\beta$ & $[0.05,\,4]$ & $[0.05,\,4]$ & $[-0.94,\,0.94]$\tablenotemark{a}
\enddata
\tablenotetext{a}{The ranges of the exponents allow at most a factor of 2.6 change between $z=2$ and $z=7$. In addition, $0\le\alpha(z)\le4$ and $0.05\le\beta(z)\le4$ at every $0\le z\le7$.}
\end{deluxetable}

\subsubsection{The bound on the low-mass slope} \label{sec:alpha_prior}

The data used here do not measure $\alpha$. Every constraint is an integral over halo mass (Equations~\ref{eq:Tb} and \ref{eq:LF}), and once $\alpha$ is large enough that halos below $\Mone$ contribute little to those integrals a further increase changes nothing. The likelihood is therefore flat along a direction that extends to arbitrarily sharp cut-offs, and a bound is needed. Every fit in this paper requires $\alpha\le4$ at every redshift.

This upper bound on $\alpha$ rests on a physical argument for how steep the relation could be. The SFR inside a halo can be modeled as its gas accretion rate times a star-formation efficiency, ${\rm SFR}=\epsilon\,\dot M_h$. Below $10^{11.5}\,\Msun$ both factors rise with halo mass: the efficiency measured from the clustering of Lyman-break galaxies at $z\simeq2$--7 scales as $\epsilon\propto\Mh^{1.2}$ below its peak \citep{Harikane2022}, and the accretion rate as $\dot M_h\propto\Mh^{1.1}$, so ${\rm SFR}\propto\Mh^{2.3}$. \cii\ in turn traces star formation nearly linearly, $\LCII\propto{\rm SFR}^{1.0-1.2}$ \citep{DeLooze2014, Schaerer2020}. Together these give $L\propto\Mh^{2.3-2.8}$, or $\alpha\simeq1.3$--1.8. The semi-analytic model of \citet{Yang2022} gives $\alpha=1.7$--2.0 at $z=1$--9. A speculative \cii\ deficit from low-metallicity galaxies would steepen the relation below $\Mone$, so the bound is placed at twice the steepest of these estimates. Section~\ref{sec:pk} shows that the power spectrum depends little on those halos, which is itself one of the results of this work.

\section{Results and discussion} \label{sec:results}

The fits of this section rest on the data of Table~\ref{tab:data} as they stand. Each result is presented with the data that drive it and compared with a joint fit with an assumed evolution of the double power-law parameters. Section~\ref{sec:implications} draws the consequences for a LIM experiment and identifies the measurement that would move the calibration forward.

\begin{figure*}
\centering
\includegraphics[width=0.8\textwidth]{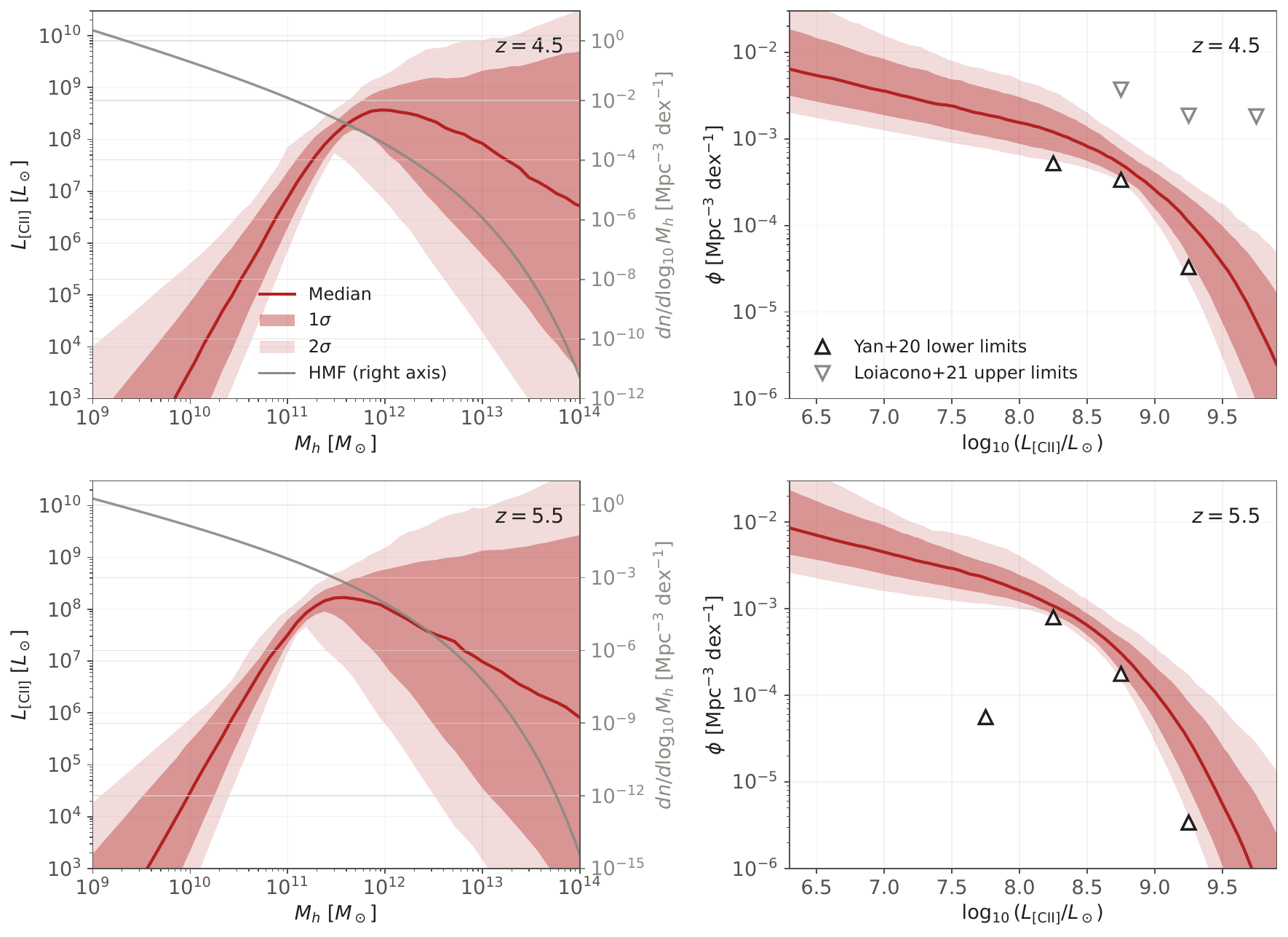}
\caption{Single-redshift fits at $z=4.5$ (top) and $z=5.5$ (bottom). Left: $L(\Mh)$. The gray curve is the HMF $dn/d\log_{10}\Mh$ at the same redshift (right axis). Right: LF with the data. Upward triangles are the ALPINE lower limits of \citet{Yan2020} and gray downward triangles the upper limits from the serendipitous sources of \citet{Loiacono2021}, each at the position the fit uses. Bands are the central 68\,\% ($1\sigma$) and 95\,\% ($2\sigma$) of the posterior and the line is its median. \label{fig:singlez}}
\end{figure*}

\subsection{Fits at a single redshift} \label{sec:singlez}

\begin{figure*}
\centering
\includegraphics[width=0.85\textwidth]{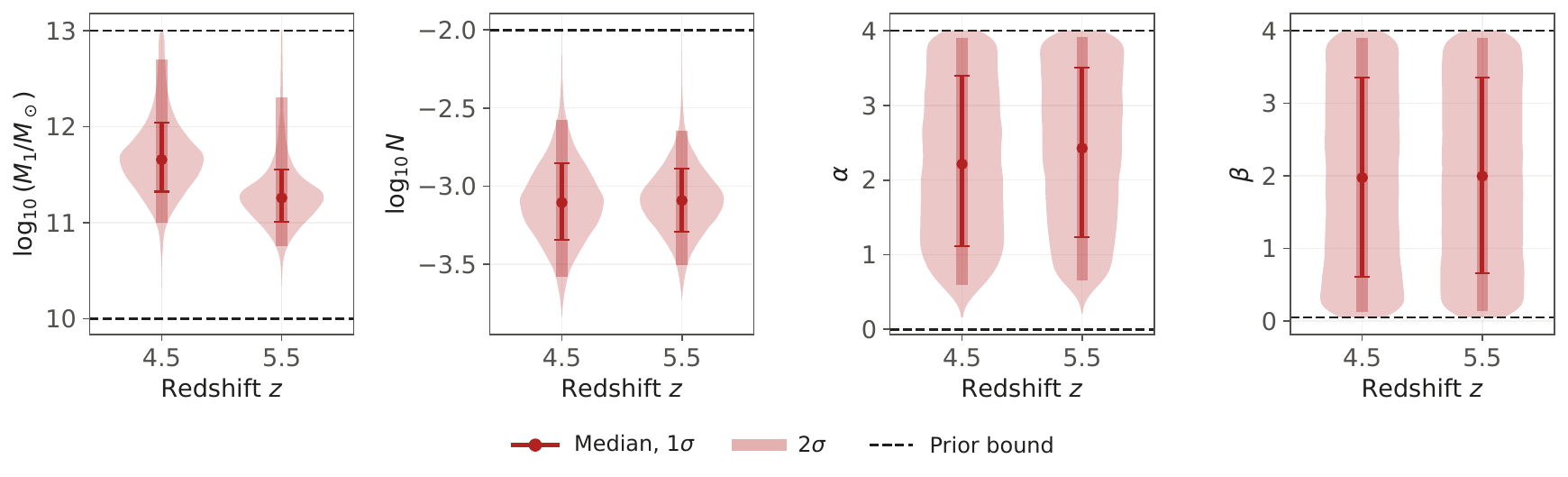}
\caption{Posteriors of the four parameters of the single-redshift fits at $z=4.5$ and $z=5.5$. Violins show the full posterior and dashed black lines the edges of the prior range. \label{fig:singlez_params}}
\end{figure*}

At each of $z=4.5$ and $z=5.5$ the four parameters of Equation~(\ref{eq:LM}) are fitted to the $\Tb$, ALPINE, and serendipitous LF constraints at that redshift alone, with no assumption about their evolution. Figure~\ref{fig:singlez} shows the resulting $L(\Mh)$ and LF, and Table~\ref{tab:singlez} and Figure~\ref{fig:singlez_params} give the parameters.

\begin{deluxetable}{lcc}
\tablecaption{Single-redshift fits \label{tab:singlez}}
\tablehead{\colhead{Quantity} & \colhead{$z=4.5$} & \colhead{$z=5.5$}}
\startdata
$\log_{10}(\Mone/\Msun)$ & $11.66^{+0.39}_{-0.33}$ & $11.26^{+0.30}_{-0.25}$ \\
$\log_{10}N$ & $-3.11^{+0.25}_{-0.24}$ & $-3.09^{+0.20}_{-0.20}$ \\
$\alpha$ & $2.21^{+1.18}_{-1.10}$ & $2.43^{+1.08}_{-1.19}$ \\
$\beta$ & $1.97^{+1.38}_{-1.37}$ & $2.00^{+1.36}_{-1.34}$ \\
\hline
$\Tb$ [$\mu$K] & $0.168\,(0.115$--$0.227)$ & $0.121\,(0.088$--$0.163)$ \\
$\beff$ & $3.9\,(3.4$--$4.4)$ & $4.3\,(3.9$--$4.6)$ \\
$\Pshot$ [$\PkUnit$] & $11\,(5$--$39)$ & $3.8\,(2.1$--$10)$ \\
$\kc$ [$\hMpc$] & $0.37\,(0.26$--$0.50)$ & $0.44\,(0.34$--$0.55)$ \\
$\Rpf$ & $45\,(23$--$93)$ & $64\,(38$--$112)$
\enddata
\tablenotetext{}{Posterior medians and central 68\,\% intervals.}
\end{deluxetable}

The break mass and the normalization are measured at both redshifts, $\log_{10}(\Mone/\Msun)=11.66^{+0.39}_{-0.33}$ and $11.26^{+0.30}_{-0.25}$ at $z=4.5$ and 5.5, and $\log_{10}N=-3.11^{+0.25}_{-0.24}$ and $-3.09^{+0.20}_{-0.20}$. The two slopes are not. The posterior of $\beta$ is flat across its prior at both redshifts: once the brightest LF bins are satisfied, $L(\Mh)$ above the break may rise, stay flat, or fall. The posterior of $\alpha$ is broad at both redshifts and falls off only toward the shallow end of the prior (Figure~\ref{fig:singlez_params}, with correlations in Appendix~\ref{app:corner}). The ALPINE lower limits and $\Tb$ together require $L(\Mh)$ to turn off below the break, because the HMF rises steeply to low masses and a shallow relation would overshoot both, but they cannot say how sharply. The data that hold the fit are the two ALPINE bins at $10^{8.25}$ and $10^{8.75}\,\Lsun$, whose lower limits the low tail of the posterior touches at both redshifts, and the extrapolated $\Tb$. No upper limit is approached by any part of either posterior.

\begin{figure*}
\centering
\includegraphics[width=0.8\textwidth]{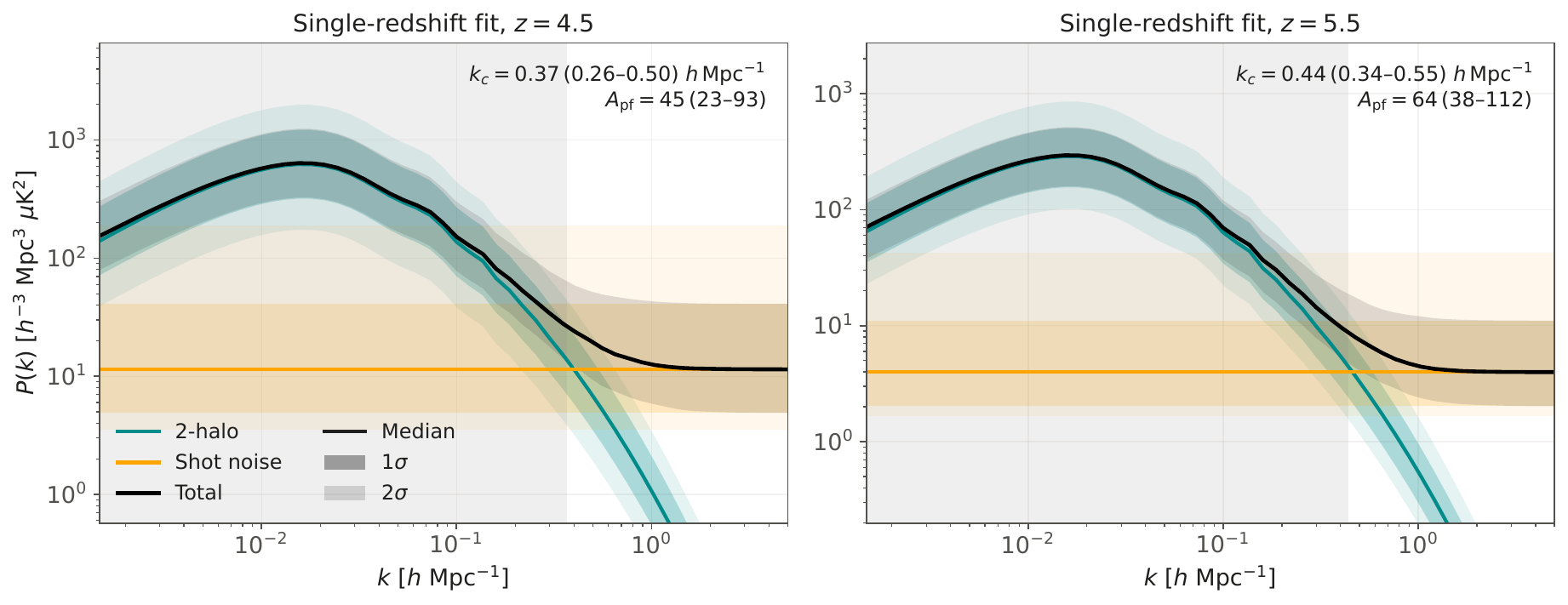}
\caption{Power spectra from the single-redshift fits at $z=4.5$ (left) and $z=5.5$ (right). Colors distinguish the two-halo term, the shot noise, and the total power, and the total is shown with its $1\sigma$ band only. The gray region marks $k<\kc$, and $\Rpf$ is the height of the two-halo peak above the shot-noise floor. \label{fig:singlez_pk}}
\end{figure*}

\subsection{Power spectrum} \label{sec:pk}

\begin{figure*}
\centering
\includegraphics[width=0.72\textwidth]{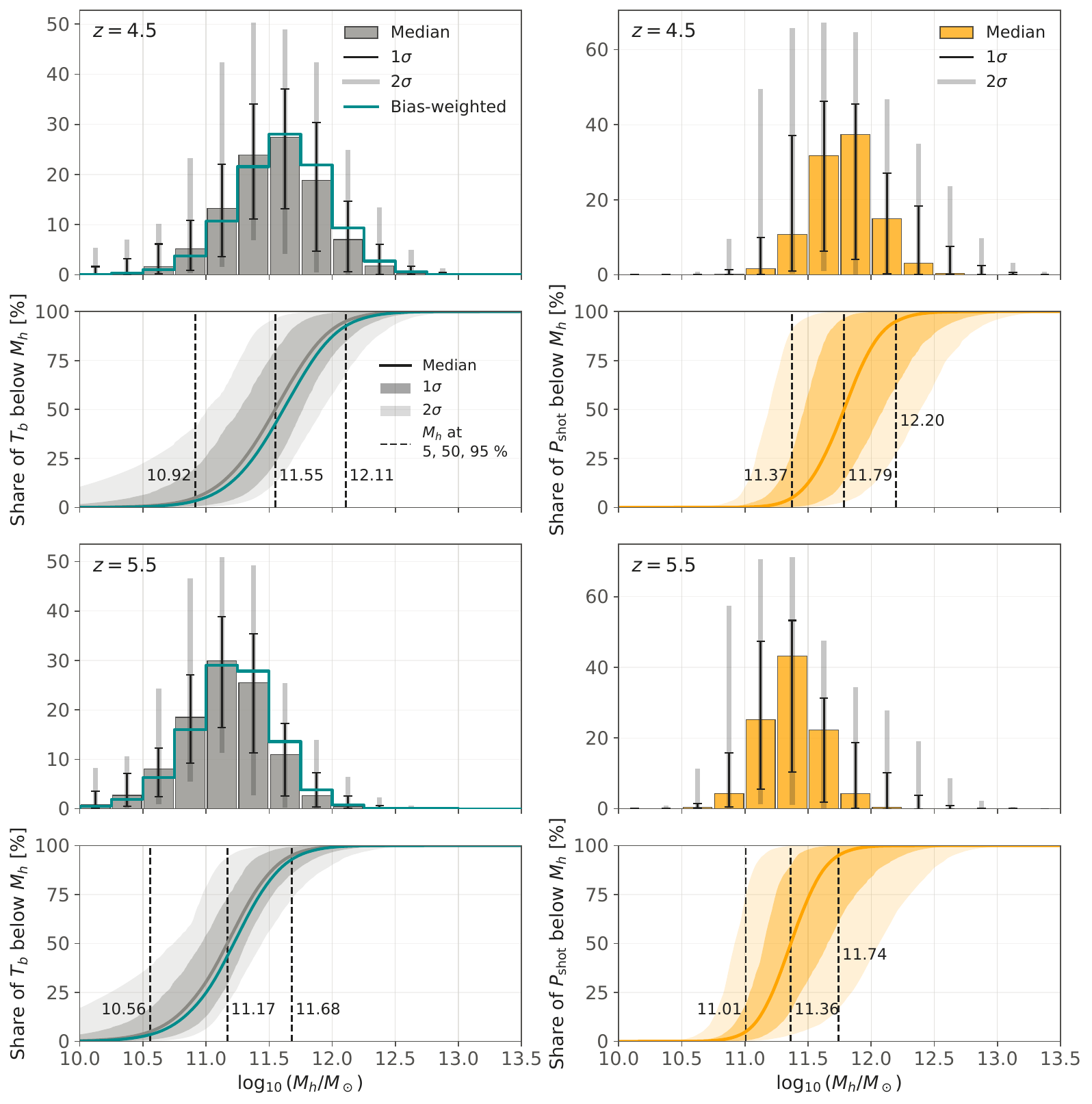}
\caption{Where the signal comes from in the single-redshift fits at $z=4.5$ (upper two rows) and $z=5.5$ (lower two rows). In the histogram rows, the bars are the fraction of $\Tb$ (left) and of $\Pshot$ (right) contributed by each 0.25\,dex bin of halo mass. The row beneath each histogram is its running total, the share of the same quantity that comes from halos lighter than $\Mh$, and the dashed lines mark the masses, written beside them as $\log_{10}(\Mh/\Msun)$, below which 5, 50, and 95\,\% of it is produced. The teal curves are the bias-weighted integrand that sets the two-halo amplitude and its running total.
\label{fig:integrand}}
\end{figure*}

Figure~\ref{fig:singlez_pk} shows the corresponding power spectra. Although two of the four parameters are unconstrained, the two-halo peak stands well above the floor in most of the posterior, $\Rpf=45\,(23$--$93)$ at $z=4.5$ and $64\,(38$--$112)$ at $z=5.5$, and it exceeds ten times the floor in 98\,\% of the samples at $z=4.5$ and in more than 99\,\% at $z=5.5$. The two-halo term and the floor rise and fall together across the posterior, because both scale with the luminosity of the same halos. The peak stands above the floor in every sample because $\Tb$ and the LF limits together pin the luminosity-weighted integrals over $10^{11}$--$10^{12}\,\Msun$, and both the clustering amplitude and the shot noise are dominated by those halos (Figure~\ref{fig:integrand}). $\alpha$ acts below that mass range, and the freedom in $\beta$ is compensated by $N$ and $\Mone$. The amplitudes are less tightly held. At $z=4.5$ the 68\,\% interval of $\Pshot$ spans a factor of 8 (Table~\ref{tab:singlez}). The shot noise is flat because each halo is a single emitter; emission extended over the halo would lower it at $k\gtrsim1\,\hMpc$ (Section~\ref{sec:observables}).

Within the assumptions of this work the luminosity-weighted bias is known to 9--12\,\%, $\beff\simeq4$ at both redshifts. This is a high bias, comparable to that of the brighter galaxies that surveys select at these redshifts. Lyman-break galaxies (LBGs) brighter than $\MUV\simeq-20.5$ have $b=3.9$--4.4 at $z=3.8$ and 5.0 at $z=4.9$ in the HSC clustering of \citet{Harikane2022}, and Ly$\alpha$ emitters (LAEs) brighter than $L^*$ have $b=4.1\pm0.2$ at $z=5.7$ \citep{Ouchi2018}, whereas the fainter LAEs of MUSE have $b=2.4$--3.1 at $z\simeq4$ \citep{HerreroAlonso2023}. The little red dots span the same range: $b=3.0\pm1.0$ at $z=3.9$--5 and $8.2\pm1.8$ at $z=5$--6 for the broad-line AGN of \citet{Lin2025}, and 3.4 and 4.5 at $z=4.5$ and 5.5 in the model of \citet{ZebrowskiNaidu2026}. Note that highly biased tracers of large-scale structure have a bias that depends on scale \citep{Kaiser1984, BBKS1986, ColeKaiser1989, MoJingWhite1997, Lazeyras2021}, and they decorrelate from the dark matter and from other tracers \citep{Wilson2019, Ravi2024, Sullivan2025, Mons2026, Ebina2026a}. The halo model used here, like other analytic treatments, includes neither effect, and models built on N-body simulations include both. A high bias would enhance the signature of local primordial non-Gaussianity, which enters as a scale-dependent bias proportional to $b-1$, but would suppress the anisotropy of the clustering from redshift-space distortions, since $f/b$ is small.  It would decorrelate the \cii\ field from other tracers, affecting the prospects for cross-correlation with galaxy surveys to control interlopers.

The shape of the power spectrum is thus settled by the existing data for the assumed scatter at fixed halo mass. Appendix~\ref{app:sigma} shows that $\Rpf$ and $\kc$ scale with that scatter, with the clustering peak above the floor in every case. The amplitude is known less well, $\Tb$ to 0.3\,dex, the power spectrum on clustering scales to 0.5--0.6\,dex, and $\Pshot$ to 0.7--0.9\,dex at 68\,\% (Table~\ref{tab:singlez}). Appendix~\ref{app:models} compares these single-redshift results with the $L(\Mh)$ models that \cii\ forecasts commonly adopt, which differ from one another by more than an order of magnitude at these redshifts.

\begin{figure}
\centering
\includegraphics[width=\linewidth]{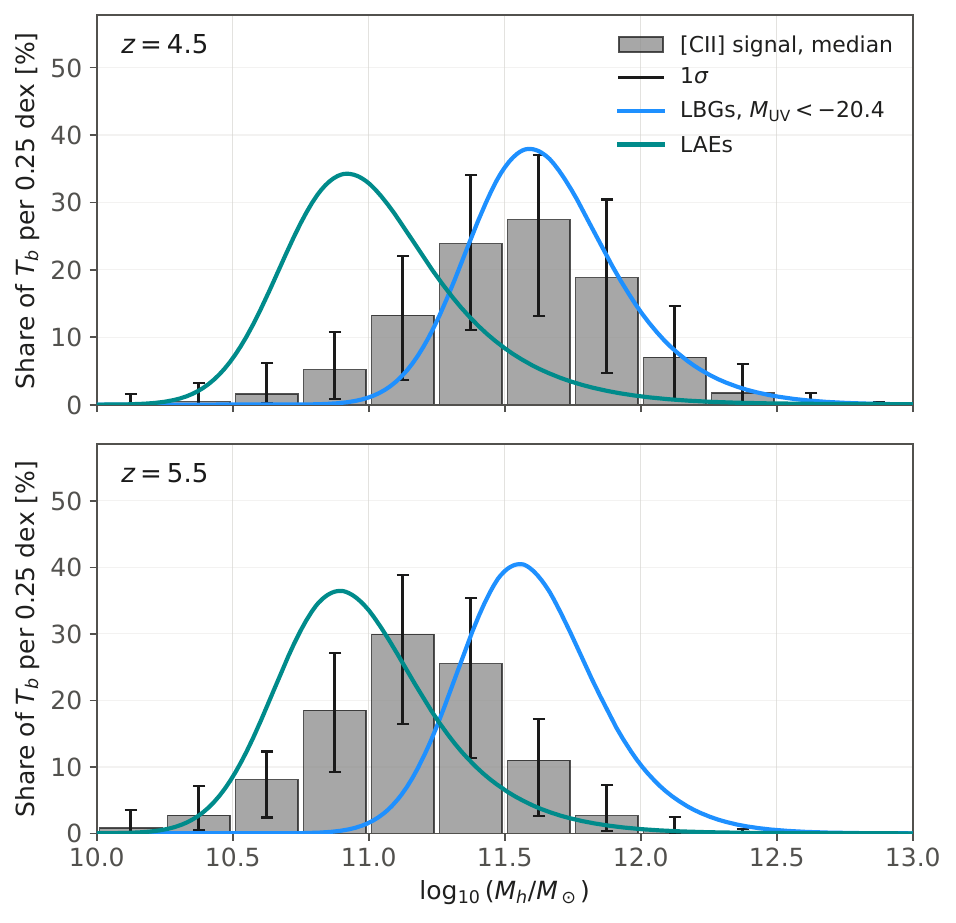}
\caption{The halo masses that produce the \cii\ signal against those of the galaxies that surveys map at similar redshifts. Bars: the share of $\Tb$ per 0.25\,dex of halo mass in the single-redshift fits at $z=4.5$ (top) and 5.5 (bottom), as in Figure~\ref{fig:integrand}. Curves: the host-halo mass distribution of LBGs brighter than $\MUV=-20.4$ from the HSC halo occupation of \citet{Harikane2022} at $z=4.9$ and of LAEs from the MUSE-Wide halo occupation of \citet{HerreroAlonso2023} at $\langle z\rangle=4.0$, each the HMF times the occupation and normalized to unit area.
\label{fig:tracers}}

\end{figure}

The robustness of the shape follows from which halos produce the signal. Figure~\ref{fig:integrand} decomposes $\Tb$ and $\Pshot$ by halo mass in the single-redshift fits. At $z=4.5$, 90\,\% of $\Tb$ comes from halos of $10^{10.9}$--$10^{12.1}\,\Msun$ and 90\,\% of $\Pshot$ from $10^{11.4}$--$10^{12.2}\,\Msun$; at $z=5.5$ the ranges are $10^{10.6}$--$10^{11.7}$ and $10^{11.0}$--$10^{11.7}\,\Msun$. The halos that dominate the clustering amplitude and those that dominate the shot noise overlap to a large degree, and the range is narrow: both are produced within about one decade of halo mass around and just above $\Mone$, which is the range where $L(\Mh)$ is best determined. The HMF sets this range (gray curve in Figure~\ref{fig:singlez}). At $z=4.5$--5.5 its knee lies below $10^{11}\,\Msun$, so $\Mone\simeq10^{11.3}$--$10^{11.7}\,\Msun$ sits on its steeply falling side, where each decade of mass holds 35--70 times fewer halos than the one below. The break mass itself follows from the knee of the LF: the ALPINE bins put it near $10^{8.5}\,\Lsun$ at a density of a few $10^{-4}$\,Mpc$^{-3}$\,dex$^{-1}$, which is the abundance of halos of $10^{11.5}$--$10^{12}\,\Msun$. Above the break there are too few halos for the behavior of $L(\Mh)$ as $\Mh\to\infty$ to matter, and below it the data require $L(\Mh)$ to fall steeply, because the far more numerous low-mass halos would otherwise overshoot $\Tb$ and the faint end of the LF. The two slopes enter accordingly. Halos below $10^{11}\,\Msun$, where $\alpha$ sets the slope, contribute $9\,(1$--$25)\,\%$ of $\Tb$ at $z=4.5$ and $31\,(13$--$55)\,\%$ at $z=5.5$, and almost none of $\Pshot$. This follows from the data, $\Tb$ together with the lower limits on the LF and the bound $\alpha\le4$, rather than from the model, and the share of the low-mass halos grows toward the shallow end of the prior, so the data do not exclude a larger contribution from faint galaxies at $z=5.5$. The high-mass slope $\beta$ enters weakly through the amplitude of the shot noise: splitting the posterior at the median of $\beta$, the half with the steeper fall-off above the break has a $\Pshot$ lower by 0.06--0.10\,dex than the other half, far less than the width of the full interval, while $\Tb$ and $\Rpf$ differ between the halves by at most 0.02\,dex. Figure~\ref{fig:tracers} places the mass distribution of the signal beside the host-halo mass distributions of the LBGs and LAEs that galaxy surveys map at these redshifts. At $z=4.5$ the signal comes from halos similar in mass to the hosts of the bright LBGs, and at $z=5.5$ it shifts toward the lighter masses of the LAE hosts, so it lies between the two distributions. The \cii\ signal therefore comes primarily from a range of halo mass that these surveys already sample. An aggregate measurement still measures the galaxies in these halos in a way that no galaxy survey does. Surveys detect such galaxies individually, but no existing spectroscopic survey covers the volumes a future LIM experiment could map at $z>4$, and the line measures the total emission of these halos, including galaxies too faint or too dusty to be cataloged.

\subsection{Joint fit} \label{sec:global}

\begin{figure}
\centering
\includegraphics[width=\linewidth]{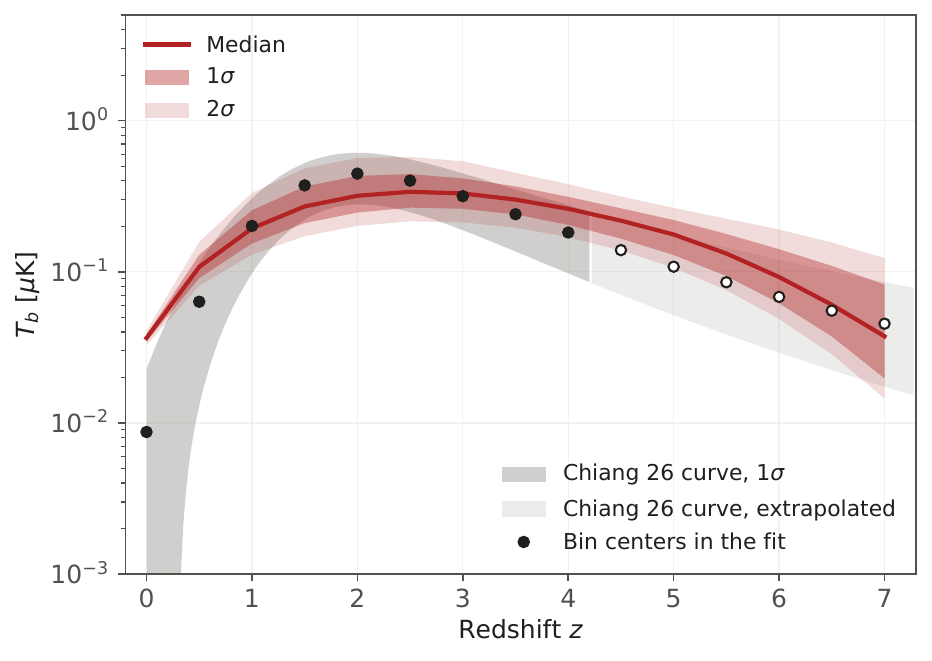}
\caption{Joint fit compared with the $T_b$ data of \citet{Chiang2026}. The gray band is the central 68\,\% of the published two-parameter curve, lighter beyond $z=4.2$ where it is an extrapolation of that form, and the black points mark its values at the bin centers that enter the likelihood (open beyond $z=4.2$), whose uncertainty is the band.
\label{fig:global_tb}}
\end{figure}

The redshift evolution of the signal is explored with the power-law model of Section~\ref{sec:evolution}. The joint fit uses all data in Table~\ref{tab:data} over $0\le z\le7$ with the evolution of Equation~(\ref{eq:zevol}). As for the single-redshift fits, every quantity quoted is a median and an interval over the posterior. Table~\ref{tab:global} lists the coefficients, Figure~\ref{fig:global_tb} compares the fit to $\Tb(z)$, and Appendix~\ref{app:joint} shows $L(\Mh)$, the LF, and the evolution of the parameters. The 14 measured numbers of Table~\ref{tab:data} are the two modes of the $\Tb$ curve and the 12 bins of the local LF. The minimum $\chi^2$ is 3.8 for these 14 numbers and eight fitted coefficients, with no limit violated, 0.02 from the $\Tb$ curve and 3.8 from the local LF. The power-law form has no justification from first principles, but it is a minimal model that is not at odds with any of the data sets.

\begin{deluxetable}{lcc}
\tablecaption{Joint fit (pivot $z_p=2$) \label{tab:global}}
\tablehead{\colhead{Parameter} & \colhead{$p1$ (value at $z=2$)} & \colhead{$p2$ (exponent of $1+z$)}}
\startdata
$\log_{10}(\Mone/\Msun)$ & $11.40^{+0.51}_{-0.24}$ & $0.20^{+1.05}_{-0.49}$ \\
$\log_{10}N$ & $-3.60^{+0.18}_{-0.14}$ & $1.24^{+0.37}_{-0.29}$ \\
$\alpha$ & $1.76^{+1.01}_{-0.62}$ & $-0.42^{+0.40}_{-0.37}$ \\
$\beta$ & $0.67^{+0.60}_{-0.33}$ & $0.04^{+0.58}_{-0.60}$
\enddata
\tablenotetext{}{Coefficients of Equation~(\ref{eq:zevol}), medians and central 68\,\% intervals.}
\end{deluxetable}

Three statements summarize the fit. First, the break mass shows no significant evolution. It is $\log_{10}(\Mone/\Msun)=11.30\,(11.23$--$11.38)$ at $z=0$, where the local LF fixes it, and its exponent, $0.20^{+1.05}_{-0.49}$, is consistent with zero, so at $z=5$ the break mass is $10^{11.4}\,\Msun$ with a 68\,\% range of $10^{11.1}$--$10^{12.3}\,\Msun$. Second, the normalization rises as $N\propto(1+z)^{1.24^{+0.37}_{-0.29}}$. Halos of fixed mass were brighter in \cii\ in the past, by a factor of $9.2\,(5.4$--$18)$ between $z=0$ and $z=5$. Third, the slopes are weakly constrained. The value of $\alpha$ at the pivot is $1.8^{+1.0}_{-0.6}$ and its evolution is consistent with zero, so the posterior spans most of the prior at every redshift. For $\beta$, the data fix the value at low redshift through the bright end of the local LF, $\beta(z=0)=0.63\,(0.59$--$0.68)$, while at $z=5$ it is $0.7\,(0.2$--$1.9)$, with a 95\,\% range of 0.1--2.9. The high-mass end of $L(\Mh)$ at $z>2$ in Figure~\ref{fig:global_evol} is therefore an extrapolation of the low-redshift LF under the assumed evolution and should be read as such.

The joint and single-redshift fits agree within their intervals, and the difference between them measures how much of the constraint at $z>4.2$ comes from the assumed evolution. At $z=4.5$ the joint fit gives $\Tb=0.216\,(0.170$--$0.266)\,\mu$K and the single-redshift fit $0.168\,(0.115$--$0.227)\,\mu$K. At $z=5.5$ they give $0.133\,(0.097$--$0.178)$ and $0.121\,(0.088$--$0.163)\,\mu$K. The joint interval is two thirds as wide in the logarithm at $z=4.5$ and the same width at $z=5.5$. The offsets of 0.11 and 0.04\,dex have a simple origin. The single-redshift fit sees only the extrapolated curve at its redshift, whose central values are 0.14 and $0.085\,\mu$K, together with the ALPINE lower limits, and it settles between them. The joint fit is anchored by the measured part of the curve at $z<4.2$ and by the local LF. Its $\Tb(z)$ is the halo model's own curve, the mass function times $N\propto(1+z)^{1.2}$, and that curve falls by 0.14\,dex between $z=3.5$ and 4.5. The published two-parameter form extrapolated beyond its last measured bin falls by 0.24\,dex over the same interval (Figure~\ref{fig:global_tb}). Beyond $z=4.2$ the joint fit therefore runs along the upper edge of the extrapolated band. The $\Tb$ offsets between the two fits at $z=4.5$ and 5.5 thus come from two different extrapolations, the halo model's and the published form's, and not from any datum at these redshifts, and both lie inside the single-redshift intervals. The single-redshift fits assume nothing about the evolution beyond the extrapolation of the measured $\Tb$ curve past $z=4.2$, and they are the headline result of this paper. The joint fit shows that the simplest evolution is consistent with every data set from $z=0$ to 7. It does not narrow the constraints at $z>4.2$, because the data there are limits only and the $(1+z)$ form leaves the exponents free, so what it adds is the continuity with the measured curve at $z<4.2$ and with the local LF. Notably, it provides a prescription for the \cii\ power spectrum at any redshift between 0 and 7, Equations~(\ref{eq:LM}) and (\ref{eq:zevol}) with the coefficients of Table~\ref{tab:global}, which future forecasts could adopt.

\begin{figure*}
\centering
\includegraphics[width=0.7\textwidth]{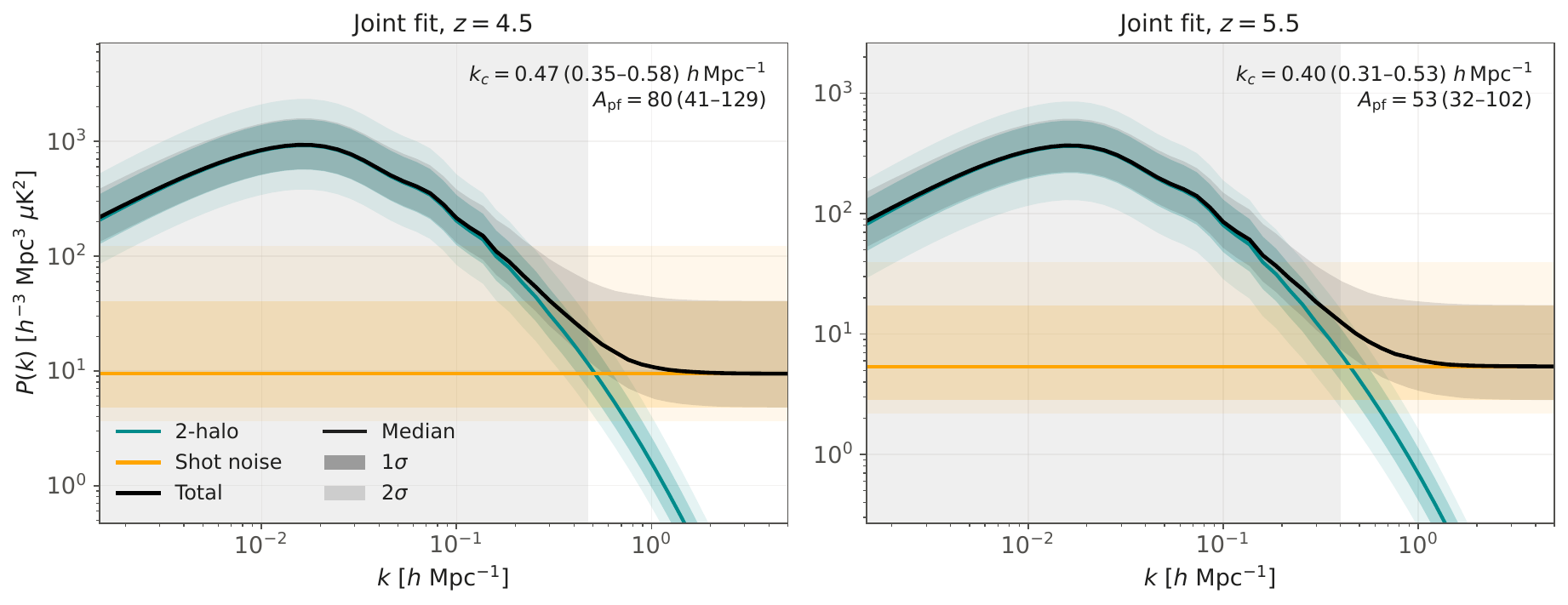}
\caption{Power spectra from the joint fit at $z=4.5$ (left) and $z=5.5$ (right). Lines and bands as in Figure~\ref{fig:singlez_pk}.
\label{fig:pk}}
\end{figure*}

\begin{deluxetable}{lcc}
\tablecaption{Power spectrum summary, joint fit \label{tab:pk}}
\tablehead{\colhead{Quantity} & \colhead{$z=4.5$} & \colhead{$z=5.5$}}
\startdata
$\Tb$ [$\mu$K] & $0.216\,(0.170$--$0.266)$ & $0.133\,(0.097$--$0.178)$ \\
$\beff$ & $3.6\,(3.3$--$3.9)$ & $4.3\,(3.9$--$4.7)$ \\
$\Pshot$ [$\PkUnit$] & $10\,(5$--$36)$ & $5.4\,(2.9$--$16)$ \\
$\kc$ [$\hMpc$] & $0.47\,(0.35$--$0.58)$ & $0.40\,(0.31$--$0.53)$ \\
$\Rpf$ & $80\,(41$--$129)$ & $53\,(32$--$102)$
\enddata
\tablenotetext{}{Medians and central 68\,\% intervals.}
\end{deluxetable}

Figure~\ref{fig:pk} and Table~\ref{tab:pk} give the power spectrum of the joint fit. The two-halo peak is $80\,(41$--$129)$ times the floor at $z=4.5$ and $53\,(32$--$102)$ times at $z=5.5$, and it exceeds ten times the floor in more than 99\,\% of the posterior samples at both redshifts. The transition from two-halo-dominated to floor-dominated scales is at $\kc=0.47\,(0.35$--$0.58)$ and $0.40\,(0.31$--$0.53)\,\hMpc$. These agree with the single-redshift values within their intervals and are no more precise. In the joint fit at $z=4.5$ the mass decomposition gives half of $\Tb$ from halos below $10^{11.3}\,\Msun$ and 90\,\% from $10^{10.3}$--$10^{12.1}\,\Msun$, with $27\,(13$--$43)\,\%$ from halos below $10^{11}\,\Msun$, the same picture as Figure~\ref{fig:integrand}.

\subsection{Implications for LIM measurements} \label{sec:implications}

Three results directly affect the design of a cosmology-focused \cii\ LIM experiment at $z>4$. First, the two-halo term exceeds the shot-noise floor at all $k<\kc\simeq0.4\,\hMpc$, and by more than an order of magnitude at its peak, in essentially all of the posterior. A measurement on these scales therefore probes the two-halo term and the underlying cosmology, not small-scale power. Second, the shape of the power spectrum is known better than its amplitude. The clustering power, $\propto\Tb^2\beff^2$, is known to a factor of 3.2--3.8 at 68\,\% and the shot noise to a factor of 5--8, whereas $\Rpf$ is known to a factor of 3--4 and $\kc$ to 30\,\%. For a noise-dominated survey the signal-to-noise ratio of the clustering amplitude scales with the clustering power, so a forecast at fixed survey parameters carries the same factor in its expected significance, while the scales on which to look for the signal and the contrast between the clustering peak and the floor are more secure. Third, the signal comes from one decade of halo mass, $10^{11}$--$10^{12}\,\Msun$ (Figure~\ref{fig:tracers}), and this identifies the measurement that would narrow the amplitude. The test is to invent a measurement at the posterior median, so that only its error bar acts, and to rerun the analysis. A deeper LF changes little. A faint point, at $10^7\,\Lsun$ to 0.15\,dex, narrows $\Tb$ by about 15\,\% and leaves $\Pshot$ unchanged, because it fixes how many galaxies have a given luminosity but not which halos host them. A bright point, at $10^9\,\Lsun$, changes nothing at $z=4.5$ and halves the $\Pshot$ interval at $z=5.5$, where the bright end is open, but a blind search would need about $10^4$ ALMA pointings to find ten such sources. What narrows the amplitude is a survey of the mean \cii\ luminosity of galaxies across the decade of halo mass that produces the signal. Points at $10^{10.9}$--$10^{12.2}\,\Msun$, each measured to 0.2\,dex, halve the $\Pshot$ interval at both redshifts and the $\Tb$ interval at $z=4.5$. Such points can come from \cii\ stacks of galaxies with spectroscopic redshifts in bins of a halo-mass proxy, for instance the UV magnitude of LBGs, whose host mass distribution is constrained from clustering \citep{Harikane2022}, or from the cross power spectrum of a \cii\ map with a galaxy sample of known bias, which measures $\Tb\beff$ directly.

\section{Conclusions} \label{sec:summary}

Working with a simplified, analytic halo model we calibrated a double power-law \cii\ halo mass--luminosity relation directly on \cii\ data: the mean brightness temperature at $0<z<4.2$ and LFs and limits at $z\simeq0$ and $4.3<z<8$. Our conclusions are as follows.

1. At $z=4.5$ and $z=5.5$, with no assumption about evolution, the break mass and the normalization of $L(\Mh)$ are measured, $\log_{10}(\Mone/\Msun)=11.66^{+0.39}_{-0.33}$ and $11.26^{+0.30}_{-0.25}$, while the low-mass slope is bounded by a physically motivated prior, $\alpha\le4$, and the high-mass slope is unconstrained.

2. The two-halo peak of the \cii\ power spectrum is $45\,(23$--$93)$ and $64\,(38$--$112)$ times the shot-noise floor at $z=4.5$ and 5.5, with $\kc=0.37$ and $0.44\,\hMpc$, and it exceeds ten times the floor in 98 and more than 99\,\% of the posterior. The amplitude is known less well, $\Tb$ to 0.3\,dex, the clustering power to 0.5--0.6\,dex, and the shot noise to 0.7--0.9\,dex.

3. Ninety percent of the mean intensity comes from halos within about one decade of mass around the break, $10^{10.9}$--$10^{12.1}\,\Msun$ at $z=4.5$ and $10^{10.6}$--$10^{11.7}\,\Msun$ at $z=5.5$, and halos below $10^{11}\,\Msun$ contribute $9\,(1$--$25)$ and $31\,(13$--$55)\,\%$. The shape of the power spectrum is therefore insensitive to the low-mass slope and to the prior on it, and it depends on the assumed scatter at fixed halo mass as Appendix~\ref{app:sigma} quantifies.

4. A joint fit of all redshifts with a single power law in $(1+z)$ per parameter is consistent with every data set from $z=0$ to 7. Within the constraining power of the data there is no evidence for a departure from that power law. In it the break mass, $\Mone\simeq10^{11.3}\,\Msun$ at $z=0$, shows no significant evolution and the normalization rises as $(1+z)^{1.2^{+0.4}_{-0.3}}$, and at $z=4.5$ and 5.5 it agrees with the single-redshift fits within their intervals. Its coefficients give a prescription for the \cii\ power spectrum at any redshift from 0 to 7.

5. A deeper luminosity function narrows the amplitude little. A measurement of the luminosity function at $10^7\,\Lsun$ to 0.15\,dex narrows the mean intensity by about 15\,\% and the shot noise not at all. One at $10^9\,\Lsun$ narrows the shot noise only at $z=5.5$, and a blind search would need about $10^4$ ALMA pointings to find ten such sources. The mean \cii\ luminosity of galaxies across the decade of halo mass that produces the signal, measured to 0.2\,dex, halves the shot-noise interval at both redshifts and the $\Tb$ interval at $z=4.5$. Stacks of galaxies with spectroscopic redshifts in bins of a halo-mass proxy, or the cross power spectrum with a galaxy sample of known bias, provide that measurement.

6. Two findings follow for experiments. The halos that produce the signal, $10^{11}$--$10^{12}\,\Msun$, are similar in mass to the hosts of the LBGs and LAEs that galaxy surveys already map at these redshifts, so \cii\ LIM probes a range of halo mass that galaxy surveys also sample. What LIM adds is the total emission of those halos, including galaxies too faint or too dusty to be cataloged, over volumes that no spectroscopic survey covers. The luminosity-weighted bias is high, $\beff\simeq4$, comparable to that of LBGs brighter than $\MUV\simeq-20.5$ and of LAEs brighter than $L^*$. This favors searches for local primordial non-Gaussianity if the large-scale systematics can be controlled, but it may also affect cross-correlations, because a highly biased field decorrelates from the underlying halos and from other tracers, an effect which is not included in our analytic model.

\section*{Data availability}
The posterior samples of the double power-law parameters from the single-redshift fits at $z=4.5$ and 5.5 and from the joint fit are publicly available on Zenodo (\url{https://zenodo.org/records/24352138}) \citep{Jeong2026data}. The derived mean brightness temperatures, luminosity-weighted biases, and shot-noise powers, along with $\Rpf$ and $\kc$, are in the same data release, at $z=4.5$ and 5.5 for the single-redshift fits and at $0\le z\le7$ in steps of 0.1 for the joint fit. The same files are available on GitHub (\url{https://github.com/objeong-astro/cii-luminosity-halo-mass}).

\begin{acknowledgments}
The authors thank Jos\'e Luis Bernal for useful discussions and for his help with the \texttt{SSLimPy} code. OJ used Claude (Anthropic) to assist with writing and debugging analysis code.
OJ was supported by CNRS/IN2P3 and by the U.S. Department of Energy.
MW was supported by the U.S. Department of Energy.
Y.-K.C. acknowledges support from the National Science and Technology Council of Taiwan (NSTC 114-2112-M-001-063-MY3) and the Academia Sinica Career Development Award (AS-CDA-113-M01).
\end{acknowledgments}

\appendix

\section{Posterior distributions} \label{app:corner}

\begin{figure*}
\centering
\includegraphics[width=0.48\textwidth]{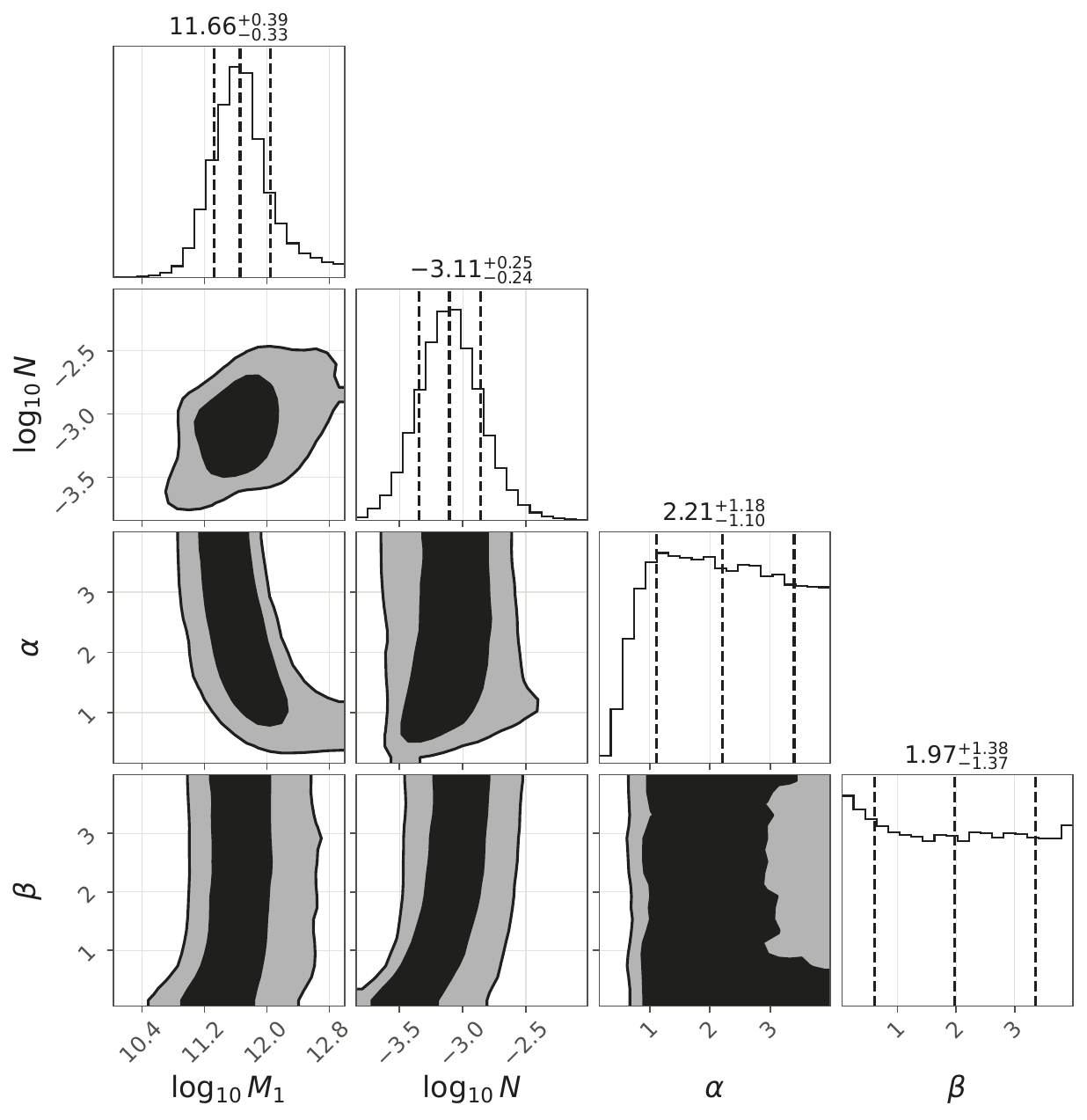}
\includegraphics[width=0.48\textwidth]{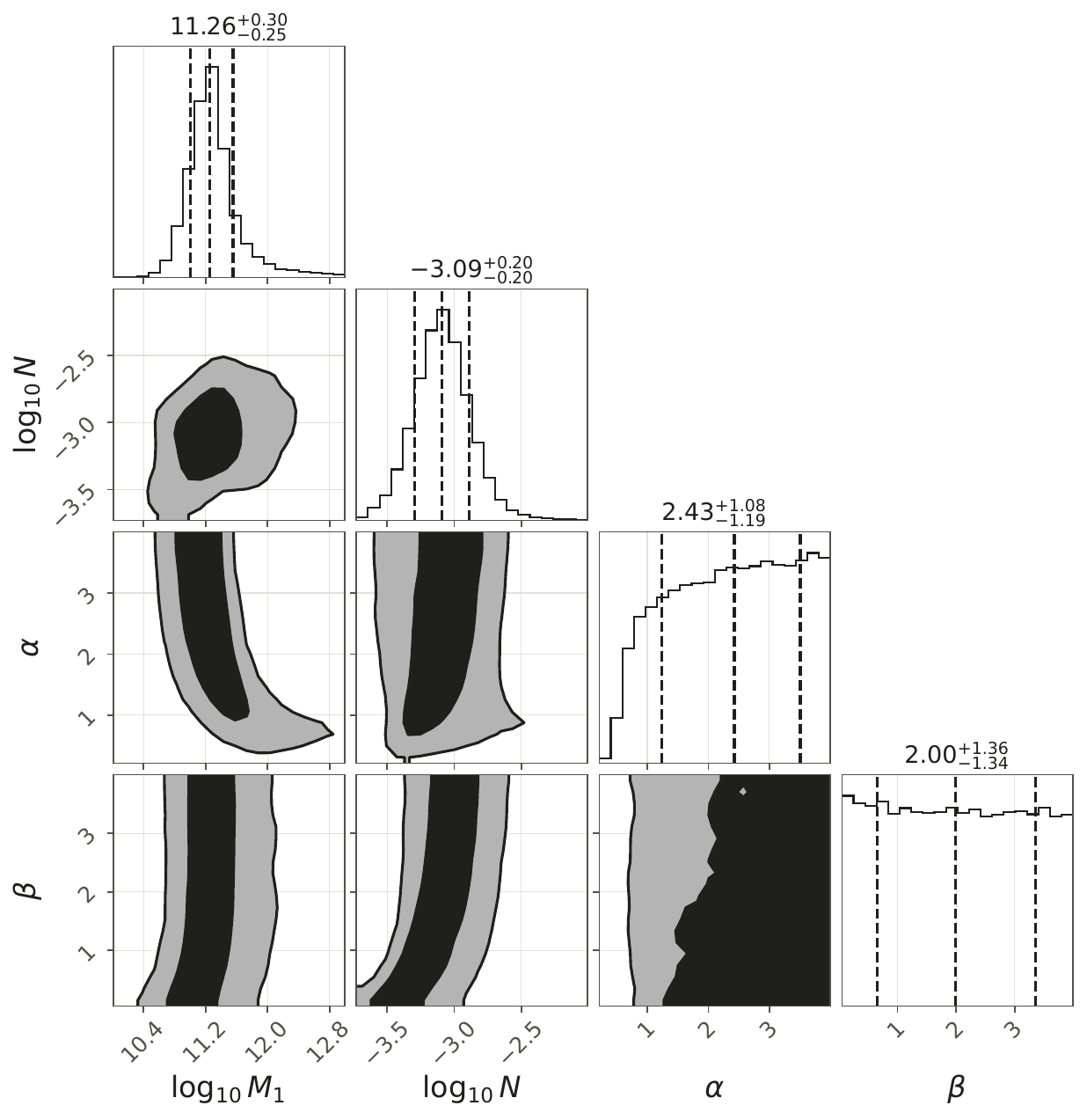}
\caption{Posteriors of the single-redshift fits at $z=4.5$ (left) and $z=5.5$ (right). Contours enclose 68\,\% and 95\,\%, and dashed lines mark the 16th, 50th, and 84th percentiles.
\label{fig:corner_singlez}}
\end{figure*}

\begin{figure*}
\centering
\includegraphics[width=0.75\textwidth]{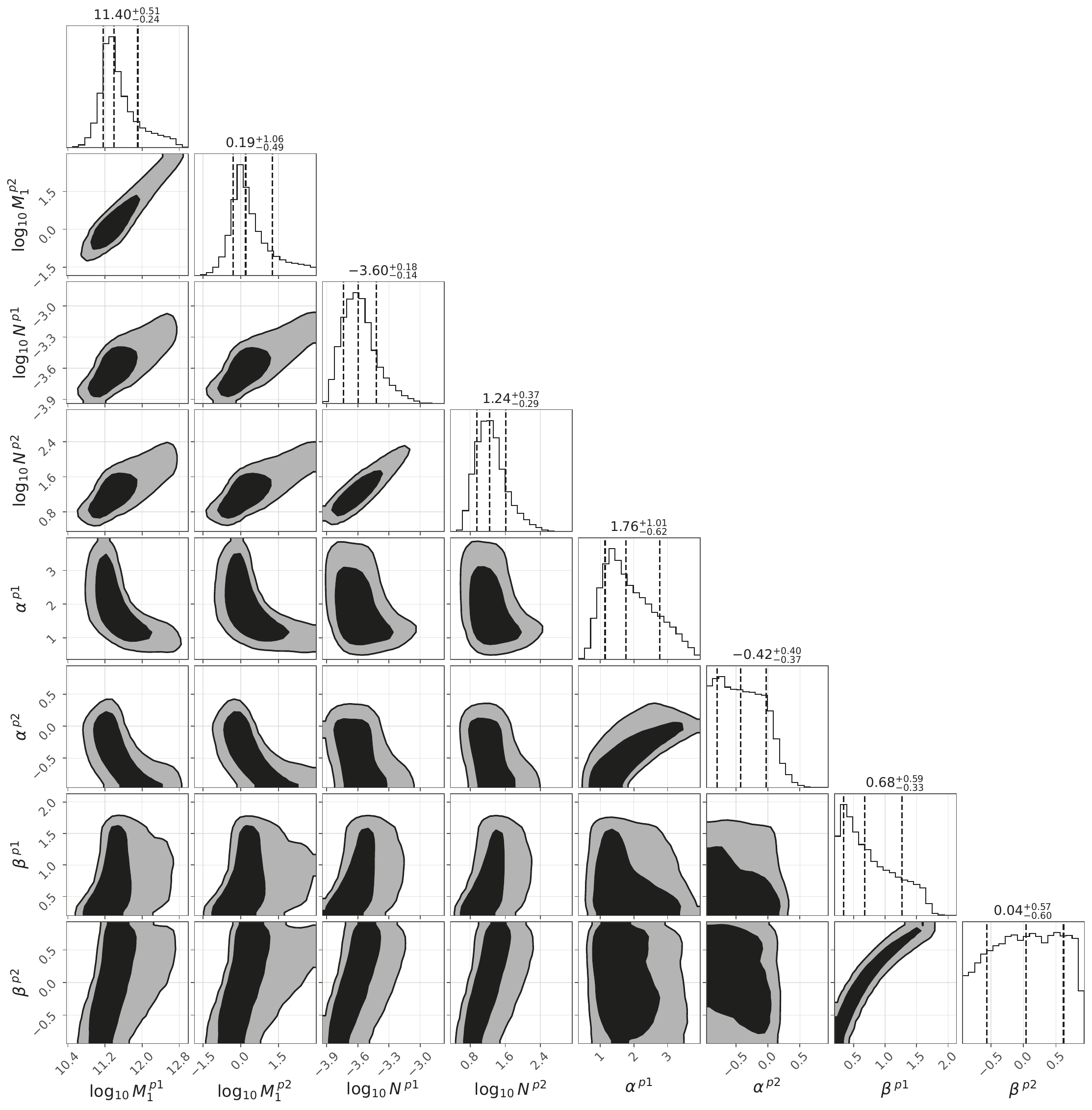}
\caption{Posterior of the eight coefficients of the joint fit. The superscripts $p1$ and $p2$ are the value at $z=2$ and the exponent of $(1+z)$. Lines as in Figure~\ref{fig:corner_singlez}.
\label{fig:corner_global}}
\end{figure*}

Figures~\ref{fig:corner_singlez} and \ref{fig:corner_global} show the posteriors of the single-redshift and joint fits.

\section{The joint fit at $z=4.5$ and $5.5$} \label{app:joint}

\begin{figure*}
\centering
\includegraphics[width=0.72\textwidth]{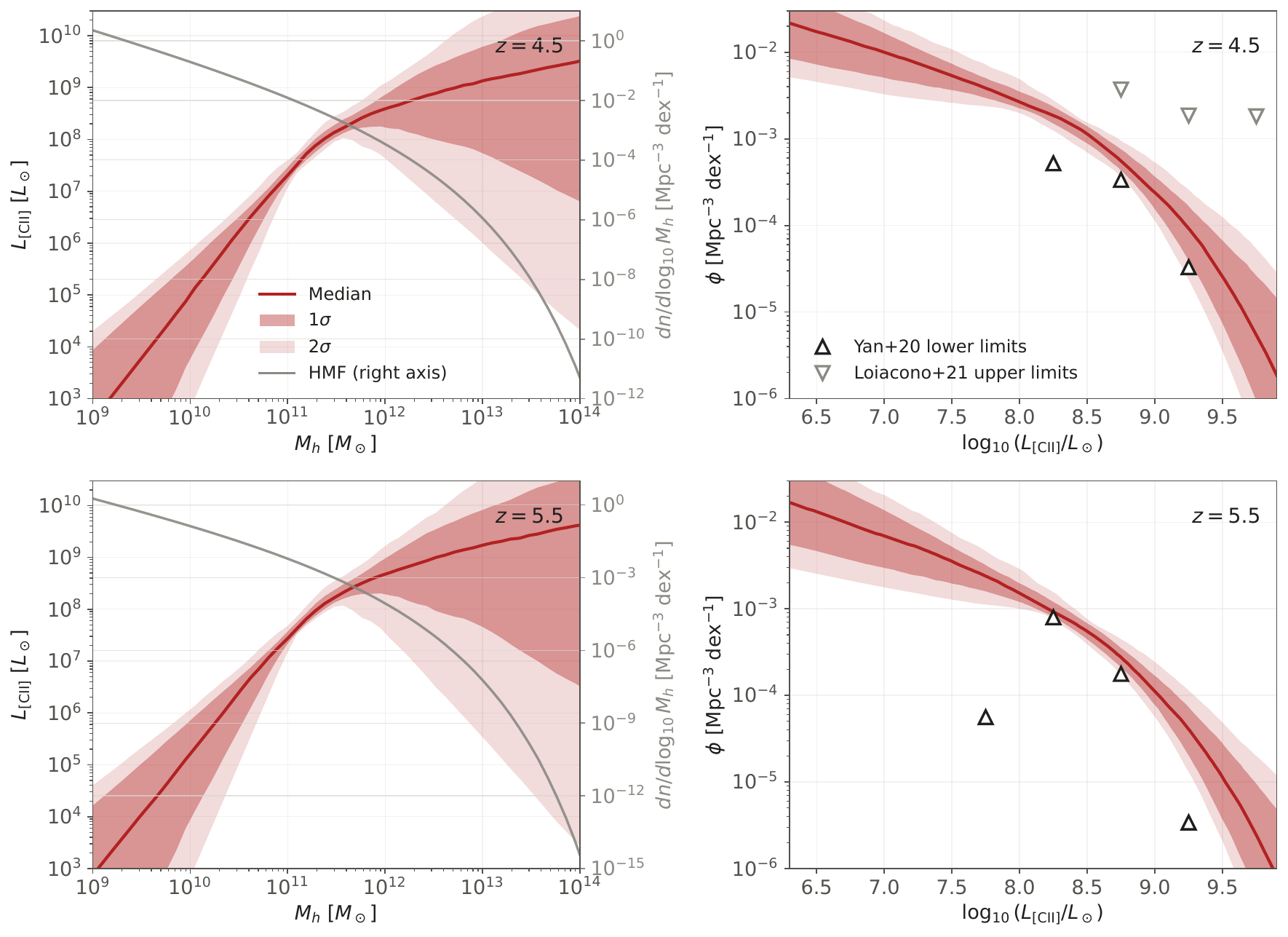}
\caption{Joint fit evaluated at $z=4.5$ (top) and $z=5.5$ (bottom): $L(\Mh)$ (left) and LF (right). Bands and symbols as in Figure~\ref{fig:singlez}.
\label{fig:global_lmlf}}
\end{figure*}

\begin{figure*}
\centering
\includegraphics[width=0.43\textwidth]{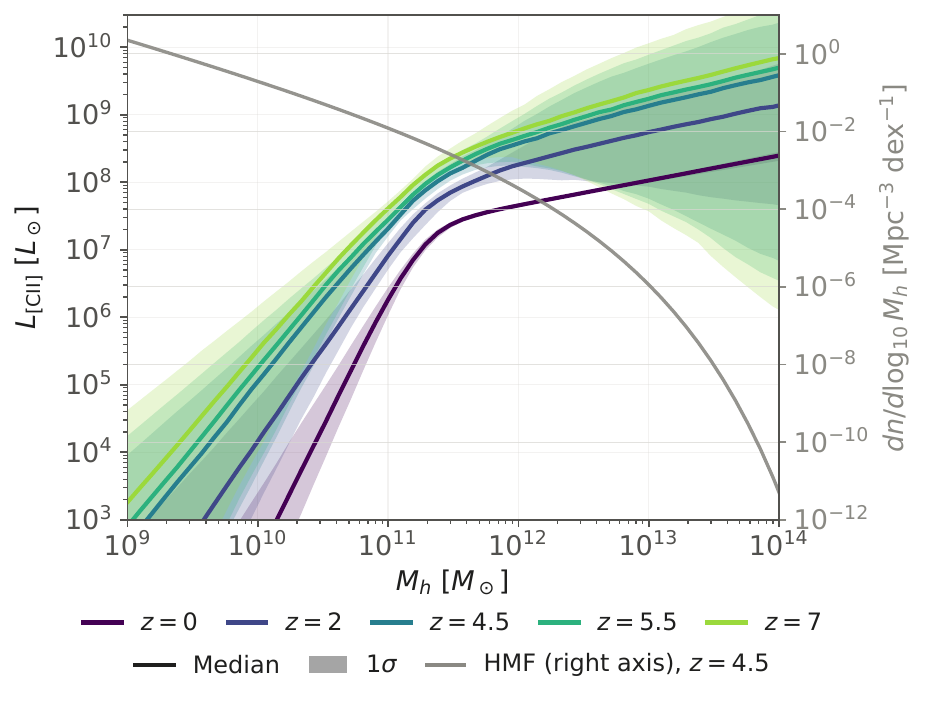}
\includegraphics[width=0.55\textwidth]{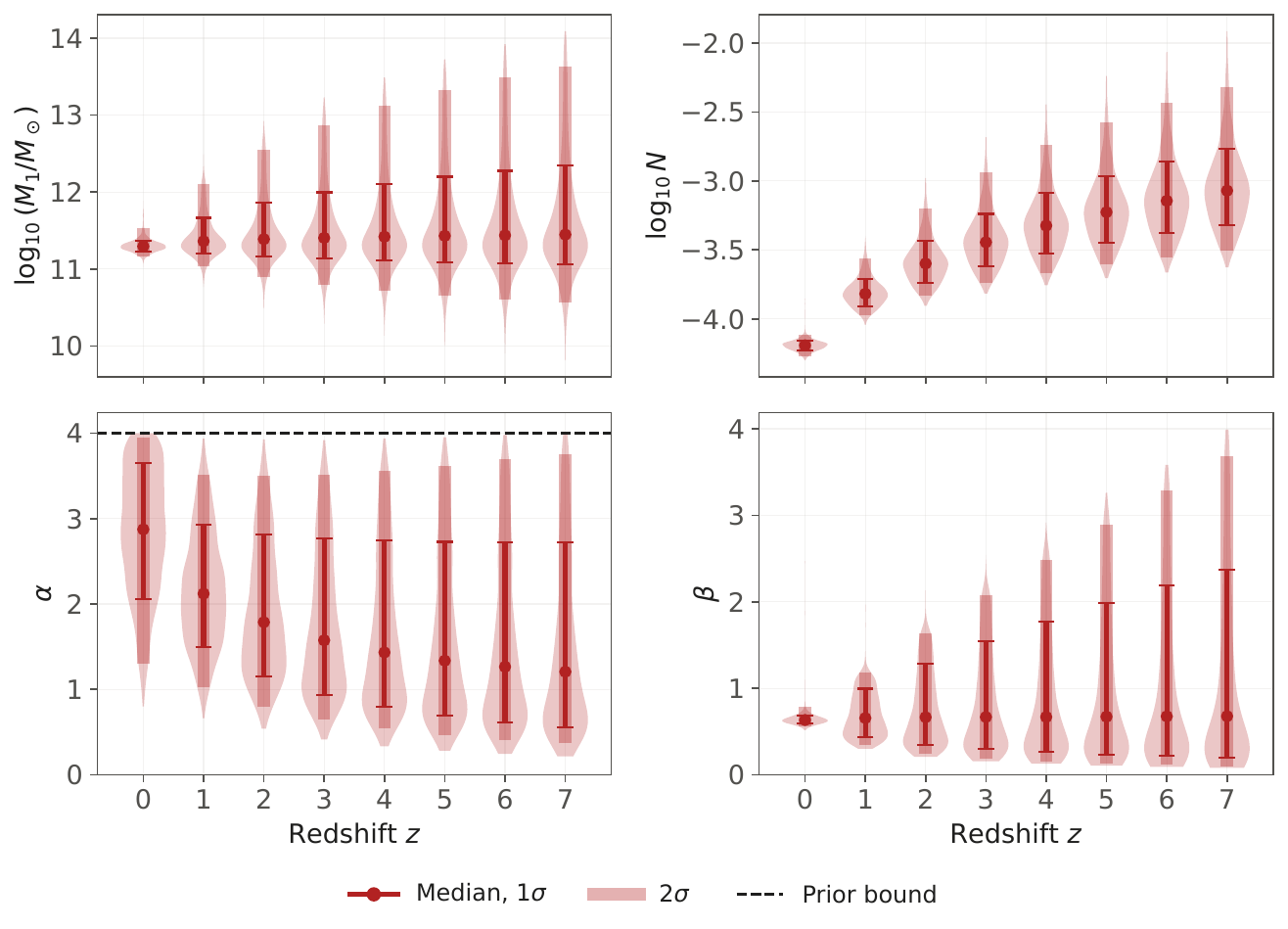}
\caption{Left: redshift evolution of $L(\Mh)$ in the joint fit. The gray curve is the HMF $dn/d\log_{10}\Mh$ at $z=4.5$ (right axis). Right: the four parameters against redshift. Violins show the full posterior and the dashed black line the prior bound on $\alpha$.
\label{fig:global_evol}}
\end{figure*}

Figures~\ref{fig:global_lmlf} and \ref{fig:global_evol} show the $L(\Mh)$ and LF of the joint fit at the two redshifts of the single-redshift fits, and the evolution of the relation and of its parameters over $0\le z\le7$.

\section{The scatter at fixed halo mass} \label{app:sigma}

\begin{figure*}
\centering
\includegraphics[width=0.8\textwidth]{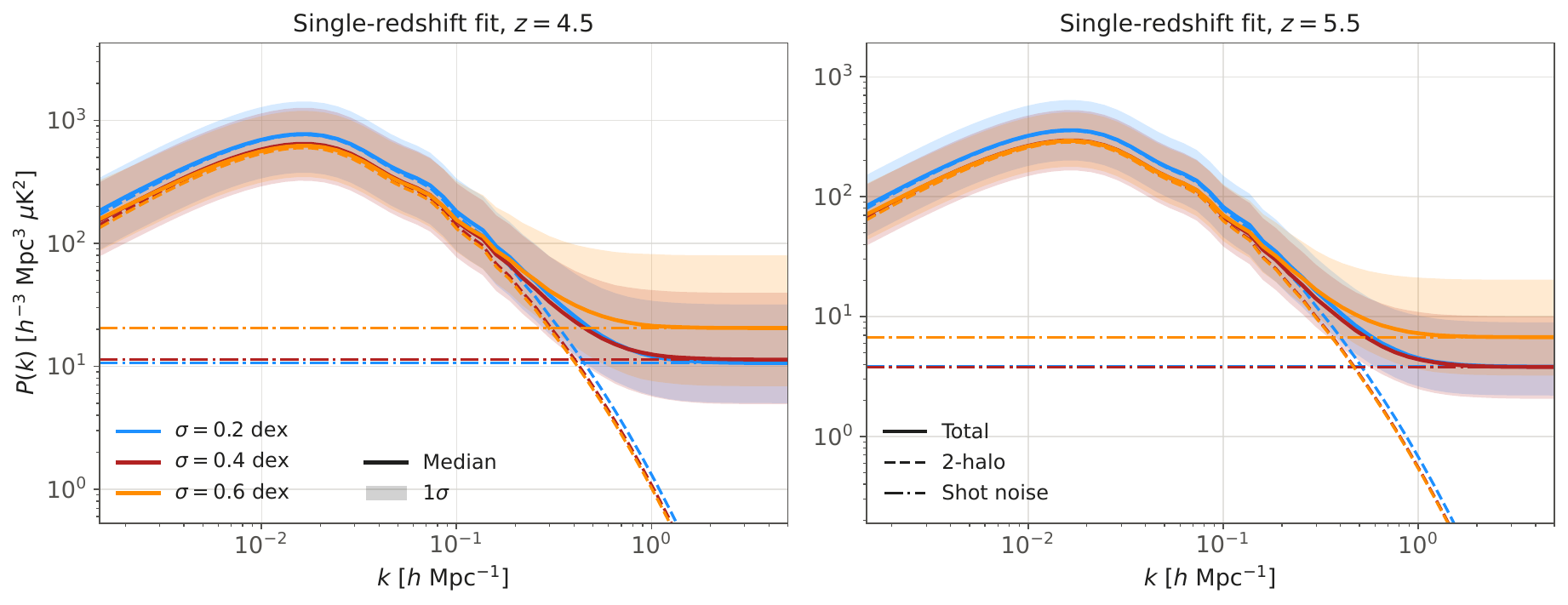}
\caption{Power spectra at $z=4.5$ (left) and $z=5.5$ (right) from the single-redshift fits with the scatter at fixed halo mass set to 0.2, 0.4 (the fiducial), and 0.6\,dex. For each scatter the total (solid, posterior median and $1\sigma$ band), the two-halo term (dashed), and the shot noise (dash-dotted) are shown.
\label{fig:sigma_pk}}
\end{figure*}

\begin{figure*}
\centering
\includegraphics[width=0.8\textwidth]{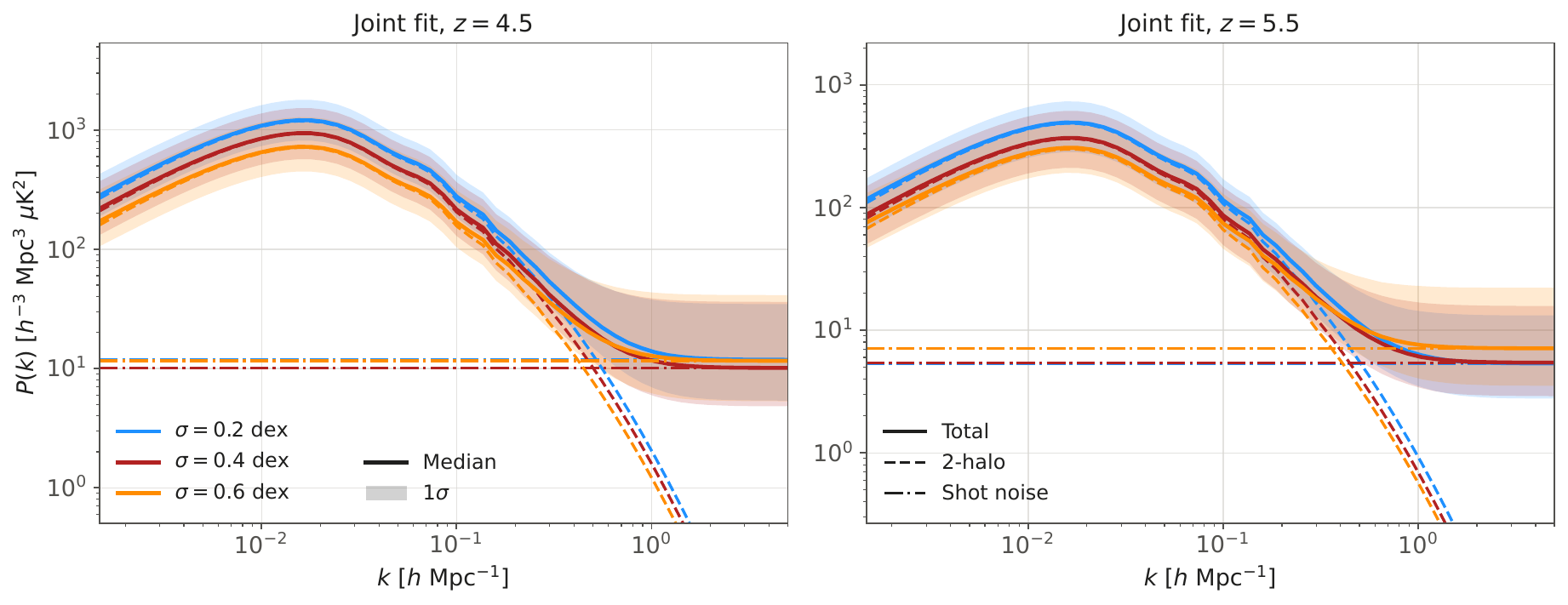}
\caption{Same as Figure~\ref{fig:sigma_pk} for the joint fit.
\label{fig:sigma_pk_joint}}
\end{figure*}

The log-normal scatter $\sigma$ of Equation~(\ref{eq:LF}) is held at 0.4\,dex in every fit. It raises both moments of the LF, the first by the factor $e^{(\sigma\ln10)^2/2}$ of Equation~(\ref{eq:Tb}), which is degenerate with $N$ and which the fit absorbs, and the second by the larger factor $e^{2(\sigma\ln10)^2}$ of Equation~(\ref{eq:Pshot}), so the data cannot determine it and its effect at fixed data falls mostly on the shot noise. To show how much of the shape of the power spectrum rests on the choice, the joint fit and the single-redshift fits at $z=4.5$ and 5.5 were repeated with $\sigma$ fixed at 0.2 and at 0.6\,dex, everything else unchanged. Figures~\ref{fig:sigma_pk} and \ref{fig:sigma_pk_joint} show the power spectra and Table~\ref{tab:sigma} the summary numbers. The scatter changes the shape of the power spectrum, not only its amplitude. At fixed data the fit readjusts $L(\Mh)$ to the same counts and the same $\Tb$, so $\Tb$ moves little. A wider scatter lets the more numerous lighter halos reach the luminosities of the LF bins, and the fit moves the break mass down, to $\log_{10}(\Mone/\Msun)=11.8$, 11.7, and 11.5 in the single-redshift fit at $z=4.5$ for $\sigma=0.2$, 0.4, and 0.6\,dex. The signal then comes from less biased halos, and $\beff$ falls from 4.1 to 3.9 to 3.6. The shot noise, which follows the second moment of the LF, stays near 11\,$\PkUnit$ up to 0.4\,dex and almost doubles at 0.6\,dex. Both effects lower the clustering peak against the floor, so $\Rpf$ falls from 58 to 45 to 26 and $\kc$ from 0.41 to 0.37 to 0.28\,$\hMpc$ (Table~\ref{tab:sigma}). At $z=5.5$ the same sequence is $\Rpf=79$, 64, and 36 and $\kc=0.48$, 0.44, and 0.33\,$\hMpc$. The joint fit follows the same trend. The two-halo term exceeds ten times the floor at its peak in at least 84\,\% of the posterior for every scatter, so the conclusion that a measurement at $k<\kc$ is dominated by clustering holds for any $\sigma$ in this range, but the values of $\Rpf$ and $\kc$ quoted in Section~\ref{sec:pk} are those of the fiducial 0.4\,dex. The data are not indifferent to the choice. The local LF of \citet{Hemmati2017} measures the bright end at $z\simeq0$, whose fall-off is set by the scatter as much as by $\beta$, and because $\sigma$ is one number at all redshifts that constraint reaches the joint fit everywhere. The minimum $\chi^2$ of the joint fit is 2.8, 3.8, and 20.1 for $\sigma=0.2$, 0.4, and 0.6\,dex, all of the difference from the local LF. A scatter of 0.6\,dex is therefore disfavored by $\Delta\chi^2=16$, while 0.2 and 0.4\,dex are equally acceptable. The single-redshift fits, which have no local LF, reach $\chi^2=0$ for every scatter and cannot tell them apart. The fiducial 0.4\,dex lies inside the range the local LF allows, and a measurement of the scatter at $z>4$, the width of the distribution of $\LCII$ at fixed halo mass, is what would fix $\Rpf$ and $\kc$ beyond the fiducial assumption.

\begin{table*}
\centering
\caption{Power-spectrum summary against the scatter at fixed halo mass \label{tab:sigma}}
\begin{tabular}{llccccc}
\hline\hline
Fit & $\sigma$ [dex] & $\Tb$ [$\mu$K] & $\beff$ & $\Pshot$ [$\PkUnit$] & $\kc$ [$\hMpc$] & $\Rpf$ \\
\hline
Single-redshift fit, $z = 4.5$ & 0.2 & $0.171\,(0.118$--$0.232)$ & $4.14\,(3.74$--$4.55)$ & $11\,(5$--$32)$ & $0.41\,(0.32$--$0.53)$ & $58\,(34$--$104)$ \\
 & 0.4 & $0.168\,(0.115$--$0.227)$ & $3.87\,(3.41$--$4.38)$ & $11\,(5$--$39)$ & $0.37\,(0.26$--$0.50)$ & $45\,(23$--$93)$ \\
 & 0.6 & $0.177\,(0.130$--$0.232)$ & $3.55\,(3.01$--$4.15)$ & $20\,(7$--$80)$ & $0.28\,(0.18$--$0.43)$ & $26\,(11$--$65)$ \\
Single-redshift fit, $z = 5.5$ & 0.2 & $0.126\,(0.092$--$0.168)$ & $4.58\,(4.27$--$4.86)$ & $4\,(2$--$9)$ & $0.48\,(0.41$--$0.57)$ & $79\,(56$--$121)$ \\
 & 0.4 & $0.121\,(0.088$--$0.163)$ & $4.29\,(3.90$--$4.65)$ & $4\,(2$--$10)$ & $0.44\,(0.34$--$0.55)$ & $64\,(38$--$112)$ \\
 & 0.6 & $0.128\,(0.099$--$0.168)$ & $3.96\,(3.54$--$4.38)$ & $7\,(3$--$20)$ & $0.33\,(0.24$--$0.45)$ & $36\,(19$--$70)$ \\
Joint fit, $z = 4.5$ & 0.2 & $0.235\,(0.193$--$0.283)$ & $3.74\,(3.43$--$4.04)$ & $12\,(5$--$35)$ & $0.52\,(0.37$--$0.62)$ & $99\,(47$--$157)$ \\
 & 0.4 & $0.216\,(0.170$--$0.266)$ & $3.57\,(3.30$--$3.87)$ & $10\,(5$--$36)$ & $0.47\,(0.35$--$0.58)$ & $80\,(41$--$129)$ \\
 & 0.6 & $0.210\,(0.168$--$0.261)$ & $3.21\,(2.89$--$3.54)$ & $12\,(5$--$41)$ & $0.39\,(0.30$--$0.51)$ & $53\,(29$--$95)$ \\
Joint fit, $z = 5.5$ & 0.2 & $0.154\,(0.109$--$0.200)$ & $4.34\,(3.83$--$4.83)$ & $5\,(3$--$13)$ & $0.47\,(0.37$--$0.61)$ & $76\,(44$--$140)$ \\
 & 0.4 & $0.133\,(0.097$--$0.178)$ & $4.35\,(3.87$--$4.70)$ & $5\,(3$--$16)$ & $0.40\,(0.31$--$0.53)$ & $53\,(32$--$102)$ \\
 & 0.6 & $0.129\,(0.102$--$0.168)$ & $4.00\,(3.58$--$4.43)$ & $7\,(4$--$22)$ & $0.33\,(0.24$--$0.46)$ & $34\,(18$--$71)$ \\
\hline
\end{tabular}
\par\smallskip
{\footnotesize Medians and central 68\,\% intervals, from the chains of the fits repeated with the scatter of Equation~(\ref{eq:LF}) fixed at 0.2, 0.4 (the fiducial), and 0.6\,dex.}
\end{table*}

\section{Comparison with published models} \label{app:models}

\begin{figure*}
\centering
\includegraphics[width=\textwidth]{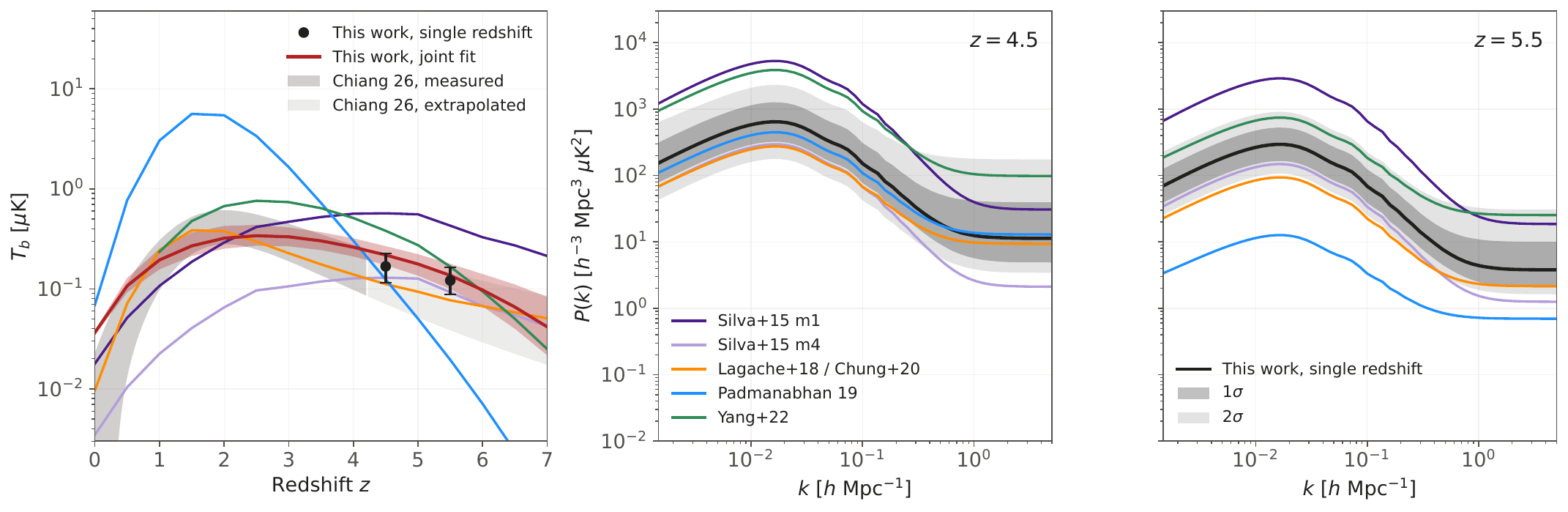}
\caption{Published $L(\Mh)$ models beside this work, all evaluated with the halo model and the 0.4\,dex scatter of this paper. Left: $\Tb(z)$ of each model against the \citet{Chiang2026} band (darker where measured, lighter where the published form is extrapolated), the joint fit of Section~\ref{sec:global} (median and $1\sigma$), and the single-redshift fits of Section~\ref{sec:singlez} at $z=4.5$ and 5.5 (points with 68\,\% intervals). The \citet{Yang2022} model is drawn over its calibrated range, $z\ge1$. Middle and right: total power spectrum of each model at $z=4.5$ and $z=5.5$ against the single-redshift fits (median, $1\sigma$, and $2\sigma$).
\label{fig:models}}
\end{figure*}

Figure~\ref{fig:models} places the single-redshift result beside $L(\Mh)$ relations that \cii\ forecasts commonly adopt. Each model is reduced to its mean $L(\Mh,z)$, with the SFR--halo mass relation it chains through where it has one, and evaluated with the HMF, bias, matter power spectrum, and scatter of this work, so the curves differ only in that relation. The models shown span the range of the published ones: the brightest and the faintest of the four $\LCII$--SFR calibrations of \citet{Silva2015} on their own SFR--halo mass relation, the redshift-dependent relation of \citet{Lagache2018} on the UniverseMachine mean SFR of \citet{Behroozi2019}, which is the fiducial model of the TIME forecast \citep{Chung2020}, the abundance-matched relation of \citet{Padmanabhan2019}, and the double power law fitted to the semi-analytic model of \citet{Yang2022} with its duty cycle.

The models span more than an order of magnitude in the total power at $k=0.1\,\hMpc$ and disagree on the shape as well as the amplitude, with $\Rpf$ from 17 to 170 and $\kc$ from 0.23 to $0.64\,\hMpc$. The mean intensity at $z<4.2$ \citep{Chiang2026}, which none of these models had, rejects most of that range (left panel). The band of this work is the range the \cii\ data allow for a double power law.

\bibliographystyle{aasjournalv7}
\bibliography{refs}

\end{document}